\documentclass{aa}

\def\v2c{v_{\rm c}^2}
\begin{document}
\input{epsf}
\newbox\grsign \setbox\grsign=\hbox{$>$} \newdimen\grdimen \grdimen=\ht\grsign
\newbox\simlessbox \newbox\simgreatbox
\setbox\simgreatbox=\hbox{\raise.5ex\hbox{$>$}\llap
     {\lower.5ex\hbox{$\sim$}}}\ht1=\grdimen\dp1=0pt
\setbox\simlessbox=\hbox{\raise.5ex\hbox{$<$}\llap
     {\lower.5ex\hbox{$\sim$}}}\ht2=\grdimen\dp2=0pt
\def\simgreat{\mathrel{\copy\simgreatbox}}
\def\simless{\mathrel{\copy\simlessbox}}
\newbox\simppropto
\setbox\simppropto=\hbox{\raise.5ex\hbox{$\sim$}\llap
     {\lower.5ex\hbox{$\propto$}}}\ht2=\grdimen\dp2=0pt
\def\simpropto{\mathrel{\copy\simppropto}}
\title{Multidimensional Hydrodynamical Simulations of radiative cooling
SNRs-clouds interactions: an application to
Starburst Environments}
\author{
Melioli C.\inst{1},
%\and
de Gouveia Dal Pino E.M.\inst{1}
\and
Raga A.\inst{2}
}
\offprints{Melioli C.}
\institute{
Universidade de S\~ao Paulo, IAG, Rua do Mat\~ao 1226,
Cidade Universit\'aria, S\~ao Paulo 05508-900, Brazil\\
 e-mail: cmelioli@astro.iag.usp.br , \ \ dalpino@astro.iag.usp.br
%\and
%Osservatorio Astronomico di Bologna, via Ranzani 1,
%Bologna, Italy\\
% e-mail: annibale@bo.astro.it
\and
Instituto de Ciencias Nucleares, Universidad Nacional Aut\'onoma de M\'exico,
Apartado Postal 70-543, 04510 M\'exico D.F., M\'exico
}

\date{Received ; accepted}

\abstract { Most galaxies present supernova shock fronts interacting with 
cloudy interstellar medium. These interactions can occur either at small 
scales, between a single supernova remnant (SNR) and a compact cloud, or at 
large scales, between a giant shell of a superbubble and a molecular cloud. 
Here study the by-products of 
SNR-clouds in a starburst (SB) system. Due to the high supernova (SN)  rate 
in this environment, a cloud may be shocked more than once by SNRs. 
These interactions can have an important role in the recycling of matter from 
the clouds to the ISM and vice-versa. Their study is also relevant to 
understand the evolution of the ISM density and the structure of the 
clouds embedded in it.

In the present work, we have focused our attention on the global effects of 
the interactions between clouds and SN shock waves in the ISM of SB 
environments and performed  three-dimensional radiative cooling 
hydrodynamical simulations  which have included the continuity equations 
for several atomic/ionic and molecular species. We have also considered the 
effects of the photo-evaporation due to the presence of the UV radiation 
from hot stars and SNe.  

The results have shown that due to the presence of 
radiative cooling, the interactions cause the formation of elongated cold 
filaments, instead of favoring an efficient  mixing of the cloud gas with 
the diffuse ISM. These filaments could  be associated with the dense clumps 
observed inside several SBs that are blown out by the galactic wind. 
The results have also revealed that the SNR-cloud interactions are less 
efficient at producing substantial mass loss from the clouds to the diffuse 
ISM  than mechanisms such as the photo-evaporation caused by  the UV flux 
from the hot stars. This result has important consequences for the global 
evolution of the SB environment and the formation of the associated 
superwinds, and may also be relevant to the ISM of normal galaxies. 

}

\titlerunning{SN heating efficiency of the ISM of SB}
\authorrunning{Melioli et al.}
\maketitle

\section{Introduction}

Starburst (SB) systems are gas-rich environments where, due to a variety of 
phenomena, the fraction of gas mass that is converted into stars, $\epsilon$, 
is $\sim$ 10 times higher than in normal galaxies (Colina et al. 1991). 
In some extreme cases, the star formation rate (SFR) can be of the order of 
100 M$_{\odot}$ yr$^{-1}$ (Scoville \& Soifer 1990), while in galaxies such 
as M82, one of the most studied SB galaxies, it is $\sim$ 5 M$_{\odot}$/yr. 
Due to the high number of massive stars, these systems present high rates of 
supernova (SN) explosions. Also, the presence of O and B stars is expected to 
produce  large UV photons fluxes that can photoionize and evaporate compact 
structures embedded in the interstellar medium (ISM) (Bertoldi 1989; 
Bertoldi \& McKee 1990; Gorti \& Hollenbach 2002).
  
Shock fronts of SN remnants (SNRs) collide with each other and with the 
clouds of the ISM, and this can lead to gas mixing and recycling of matter
from the clouds to the ISM and vice-versa. Chandra observations of SB regions 
have revealed the presence of expanding supershells (e.g. Ott, Martin \& 
Walter 2003) due to multiple SN explosions and strong stellar winds, 
therefore suggesting that interactions between shock fronts and 
inhomogeneities of the ambient medium can occur either at small scales, as 
for example, between a single SNR and a compact cloud or globule, or at large 
scales, like the encounter of a giant shell of a superbubble with a molecular 
cloud.
  
Several authors have studied shock wave-cloud interactions using both 
analytical and numerical approaches (e.g. Hartquist et al. 1986; Klein, 
McKee \& Colella 1994; Anderson et al. 1994; Jun, Jones \& Norman 1996; 
Redman, Williams \& Dyson 1998; Jun \& Jones 1999; Lim \& Raga 1999; 
de Gouveia Dal Pino 1999; Poludnenko, Frank \& Blackman 2002, hereafter 
PFB02; Fragile et al. 2004; Steffen \& Lopez 2004, Fragile et al. 2005; 
Marcolini et. al 2005), to understand the dynamics of the propagation of the 
shock front into a cloud, cloud mass loading, changes in the chemical 
composition, and the dependence of these processes on the parameters of the 
system, such as the shell and the cloud densities, the ambient temperature 
and the shell velocity. 
In general, these studies are focused on a single clump and only explore the 
processes of mass loading and the physical evolution of the shocked gas in 
the cloud. Exceptions are the studies of Steffen \& Lopez (2004), which 
investigate the interaction of a wind with several clumps, and the studies of 
Jun, Jones \& Norman (1996) and, more recently, the studies of PFB02 and 
Fragile et al. (2004), which focus on the interstellar gas enrichment that 
may result from the interaction of a planar shock wave with cylindrical 
clouds. Klein, McKee and Colella (1994), for instance, have shown that a 
cloud is destroyed in a time interval of the order of a few times the cloud 
crushing time, that is, the time needed for the internal forward shock to 
cross the cloud and reach its downstream surface. This result is confirmed by 
PFB02, and thus an efficient gas mixing is expected after the interaction of 
a shock wave with inhomogeneous ISM. However, these studies have been 
performed for adiabatic flows in a two-dimensional space. 
Since the typical timescales for radiative cooling in SBs ($10^4-10^5$ yr) 
are  smaller than their dynamical timescales, the radiative losses may have 
important effects on the dynamical evolution of these systems and therefore 
an adiabatic treatment of the flow is not completely appropriate.
 
Recently, Melioli \& de Gouveia Dal Pino (2004) have carried out a study to 
estimate the gas density evolution in SB galaxies. In such environments the 
gas exchange between the clouds and the diffuse ISM can be very fast, mainly 
due to efficient thermal and photo-evaporation of the clouds on short time 
scales compared with the dynamical time of the SB. However, these analytical 
studies have described only approximately the also important but rather 
complex interactions between these clouds and the shock fronts of the SNe. 
In the present study we try to explore the mass loss evolution of the clouds 
embedded in the ISM of a SB with the help of fully three-dimensional gas 
dynamical simulations. Employing a modified version of the YGUAZU code 
(Raga et al. 2000, Raga et al. 2002, Masciadri et al. 2002) which also 
includes the effects of gas radiative cooling and a photoionizing flux of UV 
photons due to OB stars, we investigate the evolution of the clouds that are 
shocked by SN blast waves. Though mainly focusing on the conditions of a SB 
environment, the present study also has applications to the ISM of normal 
galaxies.  

We notice that a cold cloud embedded in a hot medium could also 
evaporate at high rates due to thermal conduction effects (Begelman \& 
McKee 1990; McKee \& Begelman 1990). However,
the clouds assumed in our study must evaporate  in a time  $\sim$ 10$^6$ yr 
(Melioli \& de Gouveia Dal Pino 2004) which is much larger than the 
timescales examined here (see below) and therefore we ignored its effects. 
For a detailed study of a wind-cloud interaction including the effects of the 
thermal conduction, we refer to the recent work of Marcolini et al. (2005). 
We have also not considered the presence of magnetic fields embedded in the 
clouds and the environment.
As remarked in Melioli and de Gouveia Dal Pino (2004), they would provide 
extra support to clouds against destruction in the interactions and would 
also inhibit the effects of the thermal conduction in the direction normal 
to their dominating component. 
(Magnetic field effects will be examined in future study.)

In $\S$ 2 we discuss the numerical technique employed in the YGUAZU code and 
the input conditions of the problem. 
The results of the simulations are presented in $\S$ 3; in $\S$ 4 we 
discuss the results, then compare them with those obtained by PFB02 and 
outline the conclusions. 
Some consequences for the evolution of the ISM in SB galaxies are 
also outlined. 

\section{Numerical Simulations}
\subsection{The numerical method}

To simulate the cloud-SNR interactions we have used a modified
version of the numerical adaptive mesh refinement YGUAZUa hydrodynamical
code (see Raga et al. 2000, Raga et al. 2002) without including UV
flux, and the YGUAZUb code (Raga \& Reipurth 2004) that considers
UV flux. The YGUAZUa code integrates the gas-dynamic equations
with the "flux vector splitting" algorithm of Van Leer (1982) and
solves a system of rate equations for atomic/ionic and molecular
species. It also includes a parameterized cooling function in the
energy equation that allows the gas to cool from $\sim 10^6$ to
$1000$ K with errors smaller than 10\%. In order to integrate the system of
atomic/ionic and molecular rate equations, a semi-implicit method by
Young \& Boris (1973) is used.
The YGUAZUb code  includes the computation of the transfer of a
direct ionizing photon flux (at the Lyman limit), and the diffuse
ionizing field is considered by taking only the case B
recombination for the hydrogen.
%In this version of the code the
%radiative cooling, due to collisional excitation and
%recombination, is implicitly calculated using a time-independent
%cooling function for a gas of ISM abundances cooling from $T
%\simeq 10^6$ to $10^3$ K.

The 3-D adaptive, binary, hierarchical computational grid, in
which the gas-dynamic equations are integrated, is structured with
a base grid, and with higher resolution grids. The lower
resolution grids which correspond to $n$=1 and $n$=2, are defined
over the whole computational domain. The higher resolution grids
are defined in smaller regions of the domain and corresponds to
$n$=3,4,..,$n_{max}$, with successive increase in resolution by
factors of 2. In this study we have adopted $n_{max}$=4.

\subsection{Description of the problem and the computational domain}

Our simulations are developed in a three-dimensional computational
domain that mimics the ISM of a galaxy with the diffuse gas
characterized by an initial density  $\rho_a$, and a temperature
$T_a$, and with embedded clouds with initial density $\rho_c$, and
temperature $T_c$. These values determine the density contrast,
$\chi=\rho_c/\rho_a$, and the sound speed, $C_s^2 = \gamma k T_a /
\mu \ m_H$, where $k$ is the Boltzman constant of the gas, $\mu$
is the mean mass per particle of the gas, and $\gamma$ is the
ratio between the specific heats at constant volume and the
pressure, respectively. In this work, we take $\mu_i= 0.63$
for the ionized gas, and $\mu_n= 1.3$ for the neutral gas,
assuming a gas with 90 \% H and 10 \% He abundances,
and $\gamma$=5/3.

A shock wave front is allowed to propagate into this environment
from the bottom to the top of the box with a velocity $v_{sh} = 7
\ C_s$, so that the Mach number is  $M_s$ = 7. If the shock
front interacts with the clouds on a timescale shorter than the
cooling time of the shocked material, then the shock
front can be assumed to be  non-radiative. However,  we will
see below that this condition is not always fulfilled. From the
Rankine-Hugoniot relations, we find:

\begin{equation}
\rho_{sh} = {{2.67 M_s^2} \over {2+0.67 M_s^2}} \ \rho_g
\end{equation}
\noindent

\begin{equation}
T_{sh} = (1.25 M_s^2 - 0.25) {{\rho_g \ T_g} \over \rho_{sh}}
\end{equation}
\noindent

\begin{equation}
p_{sh} = M_s^2 p_g
\end{equation}
\noindent
where $\rho_{sh}$, $T_{sh}$ and $p_{sh}$ are the density, the
temperature and the pressure behind the shock front, respectively, and
$p_g$ is the pressure of the diffuse gas.

To resolve the physical conditions of the clouds, Klein, McKee \&
Colella (1994) have suggested that a minimum resolution of 120
cells
per cloud is necessary, but PFB02 have shown that between a
maximum and a minimum resolution of 120 and 32 cells per radius,
respectively, the only difference is the rate at which the
instabilities develop, and this does not seem to have an important
effect on the global evolution of the system. In this study, the
highest resolution grid that covers the computational volume has
1024 cells along the Z direction, 512 cells along the X direction
and 256 cells along the Y direction. This corresponds to physical
dimensions for the box of 0.8 $\times$ 0.4 $\times$ 1.6 pc,
respectively.
We notice that in some of the models to be described below,
where an otherwise unrealistically high cpu time would have
prevented us from
obtaining results within the relevant physical times of  the
interactions, we have adopted only 32 cells along the Y direction
instead of 256. In these cases, the numerical tests can be
considered  nearly 2-D, as the computational domain has an
extension of only one cloud diameter in this direction, and the
main differences with regard to previous 2-D investigations
 are the geometry of the clouds and the presence of
radiative cooling (see below).

All the clouds are assumed to be spherical with a
physical radius of 0.05 pc. The cloud diameter corresponds to 1/8
of the X-box extension and 1/16 of the Z-box extension. In the Y
direction the box is assumed to be 2 or 8 times the cloud
radius. With these dimensions, the radius of each cloud occupies
32 cells, as in the adiabatic study of PFB02. Figure 1 illustrates
the computational setup described above. We notice that numerical
tests made with a cloud occupying a number of cells that is twice
as large have produced similar results although
revealing more details in the overall filamentary structure that
develops later on in the cloud, due to the higher resolution. This
is to be expected, particularly in radiative cooling
interactions where a combination
of non-uniform radiative cooling at the shocked accelerated
regions may trigger the development of Rayleigh-Taylor
instabilities that tend to enrich the formation of
substructures, like in the shock fronts of supernova explosions and
protostellar jets (see e.g., Blondin, Fryxell and Konigl 1990; de Gouveia Dal
Pino and Benz 1993; see also the discussion in section 3.1).
However these substructures also appear in the lower resolution simulations.
On the other hand,
Kelvin-Helmholtz (K-H) instabilities, whose development would favor
fragmentation due to shear at the contact discontinuity between the clouds
and the wind, are expected to be less efficient in the presence of
radiative cooling. This tends to
produce a colder and lower pressure gas which will be less efficient
in triggering  oblique shocks and the growth of K-H unstable modes than
in adiabatic interactions (e.g., de Gouveia Dal Pino and Benz 1993,
Stone, Xu and Hardee 1997, Cerqueira and de Gouveia Dal Pino 1999).
Altogether these points suggest that the adoption in the present study of a
grid with similar resolution to that employed in earlier adiabatic studies
(PFB02, Fragile et al. 2004) will not result in the loss of relevant
information on the development of structures and fragmentation
in the SNR-cloud interaction.

\begin{figure}
\begin{center}
\epsfxsize=6cm
\epsfbox{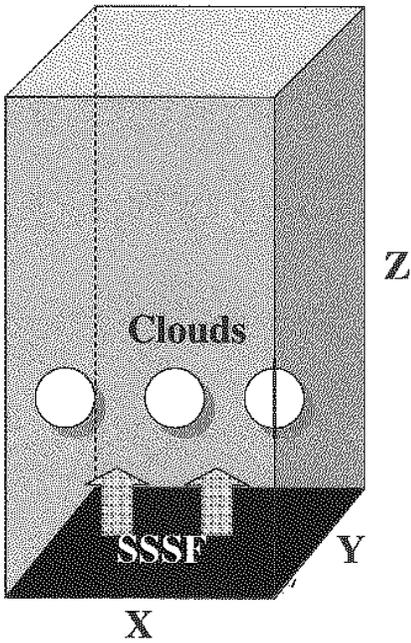}
\caption{Setup of
the computational domain (not in scale). The SSSF is injected
through the bottom boundary of the box. The boundaries are
transmitting on all the faces of the box.}
\end{center}
\end{figure}

For a better comprehension and interpretation of the SNR-cloud interactions,
we will introduce four physical parameters that are able to describe the
evolution of the system.
As in PFB02, we define the shock-crossing
time as the time necessary for the incident shock wave to sweep a single cloud:

\begin{equation}
t_{SC} = {{2r_c} \over {v_{sh}}}
\end{equation}
\noindent
where $r_c$ is the cloud radius and $v_{sh}$ is the
shell velocity. Hereafter, all the time intervals will be defined
as a function of the shock-crossing time. We note that radiative
cooling simulations are not scale-independent. In other words,
systems with different physical dimensions will result a distinct
behavior if radiative cooling is present. For this reason,
$t_{SC}$ is not the only parameter that controls the evolutionary
status of the interactions. However, $t_{SC}$ is still the best
characteristic time scale to compare the results of simulations
with different initial conditions.

We also define the normalized cloud velocity, $v_{c,N}$, as
the ratio between the velocity of the gas in the cloud core and the initial
velocity of the shock wave:

\begin{equation}
v_{c,N}={{v_{c,t}} \over {v_{sh}}}
\end{equation}
\noindent
where $v_{c,t}$ is the cloud velocity at time $t$.

Another useful parameter is the normalized mass loss, defined as
the ratio between the total mass loss of the clouds at a time $t$
($M_{l,t}$, in solar mass) and the initial total mass of the
clouds:

\begin{equation}
M_{l,n} = {{M_{l,t}} \over m_c N_c}
\end{equation}
\noindent
where $N_c$ is the number of clouds per simulation and
$m_c$ the initial mass of each cloud. This parameter gives a
measure of the effective ablation of the cloud after the passage
of a shock front at a time $t$.

We define a typical timescale
for the evolution of the cloud density. We may trace its evolution
by defining a timescale $\tau_c$ such that $\rho_c(t) =
\rho_c e^{-t/\tau_c}$, and $\tau_c$ corresponds to the time necessary to
reduce the density of the cloud by $1/e$. We will assume that all the 
gas having a density greater than $\rho_c/e$ is still part of the cloud.
This assumption will allow us
to distinguish the cloud core from the ablated gas that may eventually
mix with the diffuse ISM (see below) and therefore, to compute
the mass loss.

Finally, we define the fragmentation factor as the number of fragments,
${\cal N}_f$, per cloud, at a time $t$:

\begin{equation}
f_{f,t} = {{{\cal N}_f} \over N_c}
\end{equation}
\noindent
where we will still consider a ``cloud fragment'' as all the gas
with density greater than $1/e$ of the initial cloud density embedded 
in a more tenuous gas.

\subsection{Input conditions}

We have performed several simulations considering both adiabatic
and radiative cooling systems. In a first set, we have considered
interactions between a steady state shock front (SSSF) and a group
of clouds in the absence of a flux of UV photons. We have assumed
an initially ionized ambient medium with a mass
density $\rho_a = 2.15 \times 10^{-25}$ g cm$^{-3}$, and a temperature $T_a$ =
10$^4$ K. Clouds with mass density $\rho_c = 1.075 \times 10^{-22}$ g
cm$^{-3}$ and temperature $T_c$ = 100 K are embedded in the ambient medium.
We notice that although the clouds and the ambient medium are
initially not in perfect thermal pressure equilibrium,  this condition will 
have no
relevant effect on the interaction between the clouds and the
SSSF since the cloud sound-wave crossing time  is longer than the
dynamical timescales of the interactions. The ratio $\chi$
between the cloud and the ambient density is 500. Using
Eqs. 2 to 4, we inject an SSSF with mass density $\rho_{sh} =
8.6 \times 10^{-25}$ cm$^{-3}$ and temperature $T_{sh}= {\bf 1.5}
\times 10^5$ K with a velocity of 104 km s$^{-1}$. In this set of
simulations, only the first run (SA1) involves an adiabatic
interaction. The other simulations involve radiative-cooling
interactions of an SSSF with one (SR1), two (SR2) and three clouds (SR3).

In the second set we have performed simulations with
the same initial conditions of the first set, but assuming the presence of a
flux of UV photons which allows at least partial photo-evaporation of the
clouds.
Interactions between an SSSF with one (SRP1), two (SRP2) and
three clouds (SRP3) are also considered.

Finally, in a third set of simulations, an SNR shell (instead of a
steady state shock front) interacts with one and two clouds
without the presence of UV photons (SNS1 and SNS2, respectively),
and also in the presence of a flux of UV photons (SNSP1 and SNSP2,
respectively). In this set, the physical structure of the SNR is
similar to that described in the works of Chevelier (1974) and
Cioffi \& Shull (1992). The density and the temperature of the
ambient gas were chosen to reproduce the typical
conditions of a SB environment filled by a superbubble, with $\rho_a$ =
$2.15 \times 10^{-26}$ g cm$^{-3}$ and $T$ = 10$^4$ K.
In the cases without
photo-evaporation, we have assumed an initial density contrast between the 
clouds and the ambient medium $\chi$ = 100, for it is more appropriate
to a SB system, while in the cases with photo-evaporation, the
clouds have a much higher pressure than the ISM and the $\chi $
parameter is equal to 5000 (see,
e.g., Bertoldi and McKee 1990). The SNR shell is injected with a
velocity of 250 km s$^{-1}$ and is assumed to have a thickness
$h_{sh}$ = 1 pc, a density $\rho_{sh}= 8.6 \times 10^{-26}$ g cm$^{-3}$ and
a temperature $T_{sh} = {\bf 8.2 \times 10^5}$ K.

A summary of the parameters used in the three sets of simulations
is shown in Table 1. The simulations with the computational domain
extension in the Y direction corresponding to twice the radius of
the clouds have been labelled as 2-D to distinguish from those
(labelled 3-D) at which the box extension in this direction was
taken to be four times the cloud radius.

\begin{table*}
\caption[1]{Parameters assumed in the numerical simulations of
interactions between clouds and an SSSF or an SNR  [(1): yr; (2):
pc; (3): {2.15 $\times 10^{-24}$ g cm$^{-3}$}; (4): K; (5):
km s$^{-1}$]}
\bigskip
{\small
\centerline{
\begin{tabular}{cccccccccccccccc}
\hline\hline \noalign{\smallskip} \hbox{\small Run} & \hbox{\small
$t_{SC}^{(1)}$} & \hbox{\small UV} & \hbox{\small Cells} &
\hbox{\small Geometry} & \hbox{\small $r_c^{(2)}$} & \hbox{\small
$\rho_a^{(3)}$} & \hbox{\small $T^{(4)}$} & \hbox {\small $\rho_c^{(3)}$}
& \hbox{\small $T_c^{(4)}$} & \hbox{\small $N_c$} & \hbox{\small
$\rho_{sh}^{(3)}$} & \hbox{\small ${T_{sh}}^{(4)}$} & \hbox{\small
${v_{sh}}^{(5)}$} & \hbox{\small $h_{sh}^{(2)}$} &
\hbox{\small $M$}\\
\noalign{\smallskip}
\hline
\noalign{\smallskip}
\hbox{SA1} & \hbox{940} & \hbox{N} & \hbox{32}  & \hbox{2D} & \hbox
{0.05} &
\hbox{0.1} &
\hbox{$10^4$} & \hbox{50} & \hbox {100} & \hbox {1} & \hbox {0.4}
& \hbox {1.5 $\times 10^5$} & \hbox {104} & \hbox {$\infty$} & \hbox {7}\\
\noalign{\smallskip}
\hline
\noalign{\smallskip}
\hbox{SR1} & \hbox{940} & \hbox{N} & \hbox{32} & \hbox{3D} &
\hbox{0.05} &
\hbox{0.1} &
\hbox{$10^4$} & \hbox{50} & \hbox {100} & \hbox {1} & \hbox {0.4}
& \hbox {1.5 $\times 10^5$} & \hbox {104} & \hbox {$\infty$} & \hbox {7}\\
\noalign{\smallskip}
\hline
\noalign{\smallskip}
\hbox{SR2} & \hbox{940} & \hbox{N} & \hbox{32} & \hbox{3D} &
\hbox{0.05} &
\hbox{0.1} &
\hbox{$10^4$} & \hbox{50} & \hbox {100} & \hbox {2} & \hbox {0.4}
& \hbox {1.5 $\times 10^5$} & \hbox {104} & \hbox {$\infty$} & \hbox {7}\\
\noalign{\smallskip}
\hline
\noalign{\smallskip}
\hbox{SR3} & \hbox{940} & \hbox{N} & \hbox{32} & \hbox{3D} &
\hbox{0.05} &
\hbox{0.1} &
\hbox{$10^4$} & \hbox{50} & \hbox {100} & \hbox {3} & \hbox {0.4}
& \hbox {1.5 $\times 10^5$} & \hbox {104} & \hbox {$\infty$} & \hbox {7}\\
\noalign{\smallskip}
\hline\hline
\noalign{\smallskip}
\hbox{SRP1} & \hbox{940} & \hbox{Y} & \hbox{32} &  \hbox{3D} &
\hbox{0.05} &
\hbox{0.1} &
\hbox{$10^4$} & \hbox{50} & \hbox {100} & \hbox {1} & \hbox {0.4}
& \hbox {1.5 $\times 10^5$} & \hbox {104} & \hbox {$\infty$} & \hbox {7}\\
\noalign{\smallskip}
\hline
\noalign{\smallskip}
\hbox{SRP2} & \hbox{940} & \hbox{Y} & \hbox{32} & \hbox{2D} &
\hbox{0.05} &
\hbox{0.1} &
\hbox{$10^4$} & \hbox{50} & \hbox {100} & \hbox {2} & \hbox {0.4}
& \hbox {1.5 $\times 10^5$} & \hbox {104} & \hbox {$\infty$} & \hbox {7}\\
\noalign{\smallskip}
\hline\hline
\noalign{\smallskip}
\hbox{SNS1} &\hbox{391} &  \hbox{N} & \hbox{32} & \hbox{2D} &
\hbox{0.05} &
\hbox{0.01} &
\hbox{$10^4$} & \hbox{1} & \hbox {100} & \hbox {1} & \hbox {0.04}
& \hbox {8.2 $\times 10^5$} & \hbox {250} & \hbox {1} & \hbox {17}\\
\noalign{\smallskip}
\hline
\noalign{\smallskip}
\hbox{SNS2} & \hbox{391} & \hbox{N} & \hbox{32} & \hbox{2D} &
\hbox{0.05} &
\hbox{0.01} &
\hbox{$10^4$} & \hbox{1} & \hbox {100} & \hbox {2} & \hbox {0.04}
& \hbox {8.2 $\times 10^5$} & \hbox {250} & \hbox {1} & \hbox {17}\\
\noalign{\smallskip}
\hline
\noalign{\smallskip}
\hbox{SNSP1} & \hbox{391} & \hbox{Y} & \hbox{32} & \hbox{3D} &
\hbox{0.05} &
\hbox{0.01} &
\hbox{$10^4$} & \hbox{50} & \hbox {100} & \hbox {1} & \hbox {0.04}
& \hbox {8.2 $\times 10^5$} & \hbox {250} & \hbox {1} & \hbox {17}\\
\noalign{\smallskip}
\hline
\noalign{\smallskip}
\hbox{SNSP2} & \hbox{391} & \hbox{Y} & \hbox{32} & \hbox{2D} &
\hbox{0.05} &
\hbox{0.01} &
\hbox{$10^4$} & \hbox{50} & \hbox {100} & \hbox {2} & \hbox {0.04}
& \hbox {8.2 $\times 10^5$} & \hbox {250} & \hbox {1} & \hbox {17}\\
\noalign{\smallskip}
\hline\hline
\end{tabular}
}}
\end{table*}

\section{Results}
\subsection{Set 1}

The first set of simulations presents the interactions between
clouds and a SSSF without considering UV photon flux (runs SA1, SR1, SR2,
SR3).

When an adiabatic cloud is impacted, it passes through four characteristic
phases: compression, re-expansion, fragmentation and mixing with
the ISM (see, e.g., Nittmann et al. 1982; McKee 1988; Klein, McKee
\& Colella 1994). Its gas is initially compressed by an internal
forward shock wave that crosses the cloud in a time 
$t_{crush} \sim t_{SC} q^{0.5}$, where $q=\rho_c/\rho_{sh}$ is the
ratio between the density of the cloud and the SSSF. For the
models of the Set 1, $q=125$ \footnote {Our
value of  the density contrast between the cloud and the SSSF, $q$,
differs from that in the simulations of PFB02 by a factor of 4.}.
At the same time, a bow shock forms around the cloud and a
reverse shock is generated downstream behind the cloud. This shock
is caused by the convergence of the global flow behind the cloud
and propagates with a velocity lower than the forward internal
shock. It is dissipated very fast, compared with the
shock-crushing time, and may be considered as a sub-sonic wave
unable to change substantially the physical conditions of the
cloud. The characteristics of the initial compression phase are
well summarized in Figure 2, where we present the results of the
run SA1 for a non-radiative interaction between an SSSF and a
cloud. This result is similar to the one obtained by PFB02 for similar
initial conditions.

\begin{figure}
\begin{center}
\epsfxsize=8cm
\epsfbox{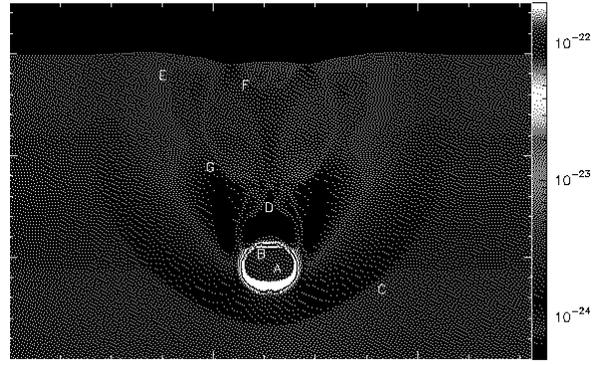}
\caption{Initial
compression phase at a time $t=3 t_{SC}$. Color-scale map of the
midplane density distribution (in log scale) for an adiabatic (or non-radiative
cooling)
interaction between an SSSF and a cloud (model SA1 of Table 1). The
density is in g cm$^{-3}$}
\end{center}
\end{figure}

As in PFB02, we can distinguish the forward (A), the reverse (B)
and the bow
shock (C), the back flow (D), the primary and secondary vortex
sheets (E, F)
and the Mach reflected shocks (G).
{\footnote {The back flow is caused by the global forward shock convergence
on the symmetry axis; the Mach reflected shocks are caused by Mach reflection
of the global forward shock at the symmetry shock; the primary and secondary
vortex sheets are caused by regular reflection of the bow shock and by the
primary Mach reflection of the forward shock, respectively (PFB02).}}
A detailed description of these structures is also presented by, e.g.,
Klein, McKee and Colella (1994) and PFB02.
The comparison of the density distribution of the system at two different
epochs (Figure 3) with the non-radiative numerical simulations of PFB02 reveals
the close similarity between both results, and therefore confirms the previous
calculations.

\begin{figure}
\center 
\epsfxsize=8cm 
\epsfbox{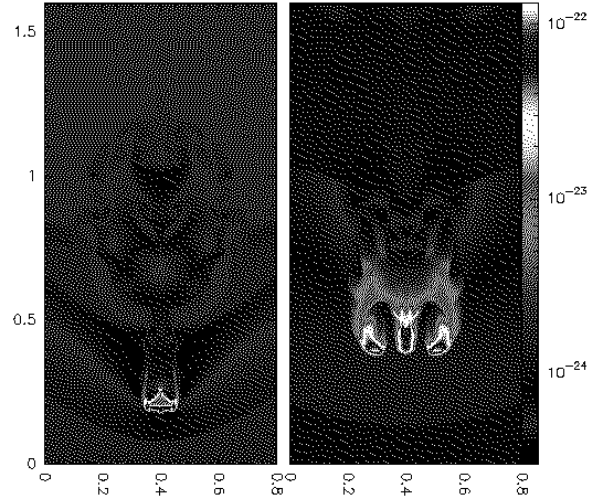} 
\caption{Color-scale
map of the midplane density distribution (in log scale) for the
non-radiative interaction of a SSSF with a cloud with $m_c \sim 0.001
M_{\odot}$, $r_c$=0.05 pc, and $\rho_c$ = 1.075 $\times 10^{-22}$ cm$^{-3}$
(model SA1 of Table 1). The shell has a density $\rho_{sh}$ = 8.6 $\times
10^{-25}$ g cm$^{-3}$ and
velocity of 104 km ${\rm s^{-1}}$, and the ambient medium has
$\rho_a$ = 2.15 $\times 10^{-25}$ g cm$^{-3}$ and T= $10^4$ K.
The computational box has
dimensions 0.8 pc $\times$ 0.8 pc $\times$ 1.6 pc, corresponding
to 512 $\times$ 512 $\times$ 1024 grid points at the highest grid
level. Time steps are $t$ = 22 $t_{SC}$ (a), and $t$ = 36 $t_{SC}$
(b). The density is shown in units of g cm$^{-3}$.}
\end{figure}

We note that after a compression phase, the cloud starts to expand
over the ambient medium. The sideways expansion of the cloud and
the relative motion between it and the SSSF excites the
development of oblique shocks and the Kelvin-Helmholtz instability
at the cloud edge that causes some entrainment of the ambient
material into the cloud. Also, the frontal acceleration of the
much denser cloud material on the more rarefied ambient gas makes
the system become Rayleigh-Taylor unstable and this causes the
development of filamentary structure or the umbrella-type
shape of the cloud seen in Figure 3. The evolution of these
instabilities causes a mixing of the cloud gas with the ISM.
Similar effects have been also found by PFB02 (see snapshots $t$ =
50.54 $t_{SC}$ and 68.40 $t_{SC}$ of their Figure 2).

On the other hand, when radiative cooling is considered, we expect
that the mixing with the interstellar gas is delayed, and this can
be seen by comparing the timescales for the setup of the
Rayleigh-Taylor (R-T) and Kelvin-Helmholtz (K-H) instabilities
with the timescale for radiative cooling.

The growth of the R-T instability due to the acceleration of the
cloud by the postshock background gas is (e.g., Fragile et al.
2004):

\begin{equation}
\tau_{R-T} \sim r_c \ {{q^{0.5}} \over {(k r_c)^{0.5} v_{sh}}}
\end{equation}
\noindent where k is the wavenumber of the unstable mode. For $q
\gg$ 1 the timescale for growth of the K-H instability is $t =
q^{0.5}/ (k v_{rel})$ (Chandrasekhar 1961), where $v_{rel}$ is the
relative velocity between the postshock background and the cloud.
Since the cloud accelerates rather slowly, the relative velocity
is approximately the same as the postshock velocity, and thus

\begin{equation}
\tau_{K-H} \sim r_c \ {{q^{0.5}} \over {(k r_c) v_{sh}}}
\end{equation}
\noindent Wavelengths corresponding to $k r_c$ =  1  are the most
disruptive, and thus $2 \tau_{R-T} \sim 2\tau_{K-H} \sim
t_{SC} \ q^{0.5} \simeq 1 \times 10^4$ yr, which is equal to the
cloud crushing time, $t_{crush}$.

Now let us estimate the radiative cooling timescales. When the
SSSF wind impacts the cloud, a  double shock structure develops at
the contact surface between them, with a forward shock that
advances into the cloud with a velocity $v_{s,c} \simeq
v_{sh}/q^{0.5}$ and a reverse shock that decelerates the SSSF wind
with a velocity $v_{s,sh}= v_{sh} - v_{s,c}$. For an SSSF
velocity  $v_{sh}= $104 km s$^{-1}$ and a density contrast between
the cloud and the SSSF of $q= 125$,
$v_{s,c} \simeq 9.3$ km $s^{-1}$ and $v_{s,sh} \simeq 95$ km
s$^{-1}$. The corresponding radiative cooling time of the shocked
gas in the cloud, for which the shock velocity is smaller than 80
km s$^{-1}$, can be estimated from Hartigan, Raymond and Hartmann
(1987; see also Gonzalez 2001) as

\begin{equation}
t_{c,c} \sim 15 \times 10^4  \ {\rm yr} \ ({{v_{s,c}} \over {9.3 \
{\rm km \ s^{-1}}}})^{-3.58} ({\rho_c \over {1.075 \times 10^{-22} \ {\rm
g \ cm^{-3}}}})^{-1}
\end{equation}
\noindent which is larger than the cloud crushing time, a though
very sensitive to the value of the shock velocity. On the other
hand, the radiative cooling time of the shocked gas behind the
SSSF, for which the shock velocity is larger than 80 km s$^{-1}$,
is

\begin{equation}
t_{c,SSSF} \sim 1.8 \times 10^4  \ {\rm yr} \ ({{v_{s,sh}} \over
{95 \ {\rm km \ s^{-1}}}})^{1.12} ({\rho_{sh} \over {8.6 \times
10^{-25} \ {\rm g \ cm^{-3}}}})^{-1}
\end{equation}

\noindent which is of the order of $t_{crush}$. As a consequence
of this rapid cooling, the shocked gas of the SSSF wind will
produce a cold thin shell around the cloud that will lessen the
damage of the impact upon it. The thickness of this cold shell
for a shock velocity larger than 80 km s$^{-1}$ is of the order of
\begin{equation}
d_{c,SSSF} \simeq 0.11 r_c  \ ({{v_{s,sh}} \over {95 \ {\rm km \
s^{-1}}}})^{4.73} ({{\rho_{sh}} \over {8.6 \times 10^{-25} \ {\rm g \
cm^{-3}}}})^{-1},
\end{equation}
which is consistent  with the thickness of the
cold material in the simulations below (see Fig. 4).

The estimates above indicate that since $t_{c,SSSF} \sim t_{crush}
\sim  2\tau_{R-T} \sim 2\tau_{K-H}$, the radiative cooling will
generally play an important role in the evolution of the clouds.
Also, as mentioned earlier, a radiative SSSF-cloud interaction
will be less efficient in producing cloud fragmentation than a
non-radiative interaction {\footnote {We also notice that
before the SSSF wind reaches the clouds, it will impact
supersonically the ambient medium at rest thus producing a shock
into it that propagates with a velocity $v_{s,a} \simeq v_{sh}/[1
+ (\rho_a/\rho_{sh})^{0.5}] \simeq 69$ (e.g., de Gouveia Dal Pino
\& Benz 1993). The shocked ambient material will also suffer
relatively rapid cooling, $t_{c,a} \sim 2.8 \times 10^4  \ {\rm
yr} \ (v_{s,a} / 69 \ {\rm km \ s^{-1}})^{-3.58} (\rho_c / 1.075
\times 10^{-22} \ {\rm g \ cm^{-3}})^{-1}$. This cold intercloud
gas will be  pushed along with the SSSF wind and also incorporate
the contact discontinuity during  the interaction of the SSSF with
the clouds.}}.

Figures 4 to 6 depict this situation.
\begin{figure}
\begin{center}
\begin{tabular}{cc}
\epsfxsize=8cm
\epsfbox{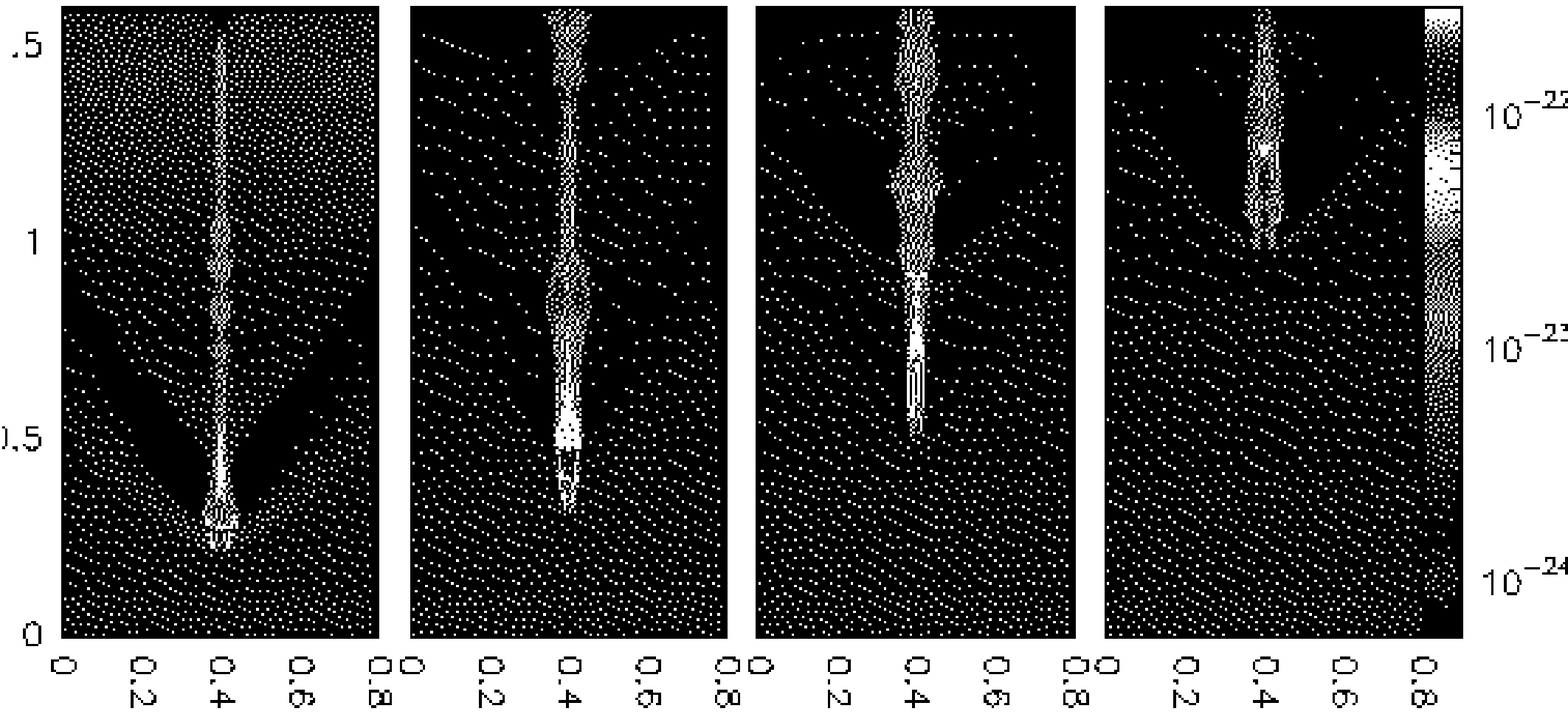}
\end{tabular}
\begin{tabular}{cc}
\epsfxsize=8cm
\epsfbox{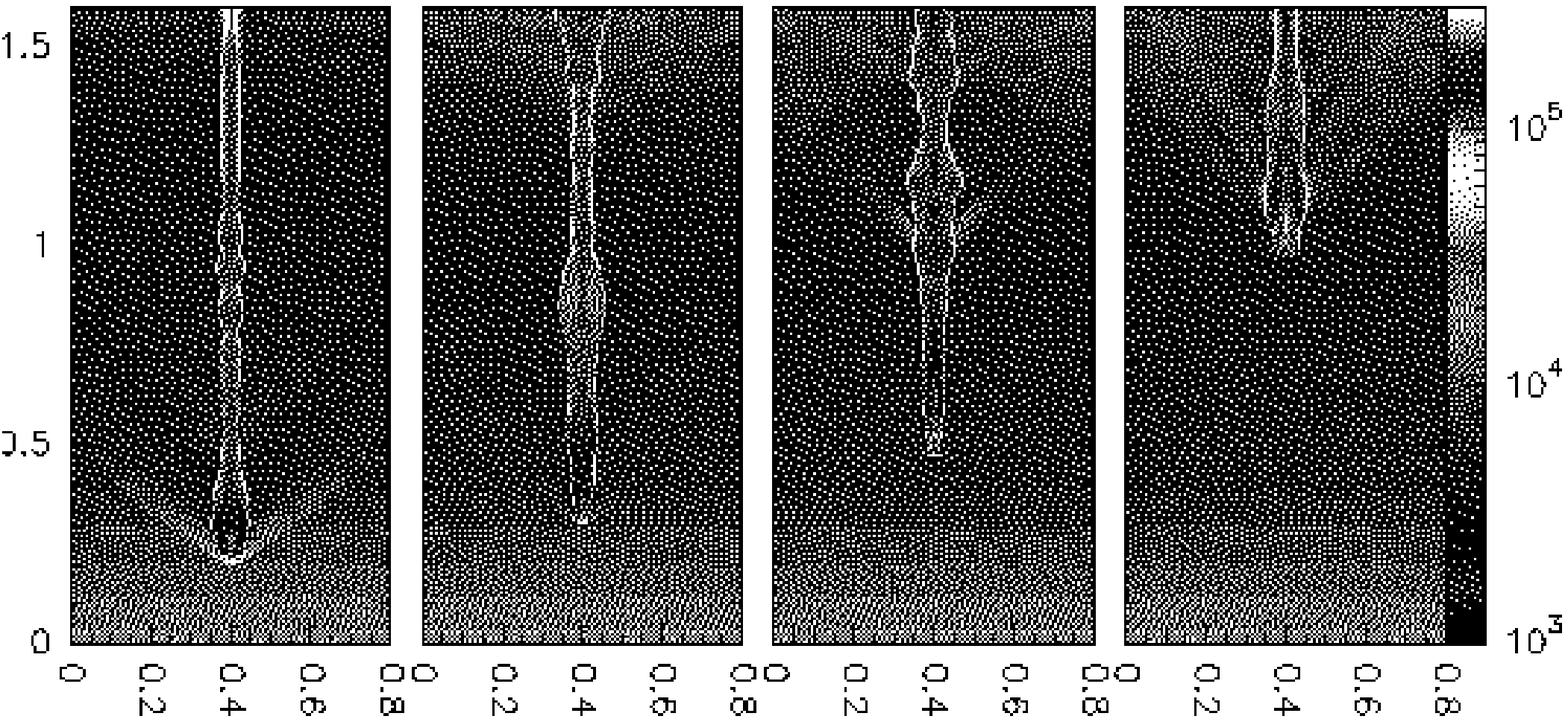}
\end{tabular}
\caption{Model SR1. Color-scale map of the midplane density (top
panel) and temperature (bottom panel) distribution (in log scale)
for the radiative cooling interaction of a SSSF with one cloud.
The initial conditions are the same as in the timescale model of
Figure 3. Time steps are $t$ = 20 $t_{SC}$ (a), $t$ = 34 $t_{SC}$
(b), $t$ = 50 $t_{SC}$ (c), and $t$ = 74 $t_{SC}$ (d).}
\end{center}
\end{figure}
\noindent
\begin{figure}
\begin{center}
\begin{tabular}{cc}
\epsfxsize=8cm
\epsfbox{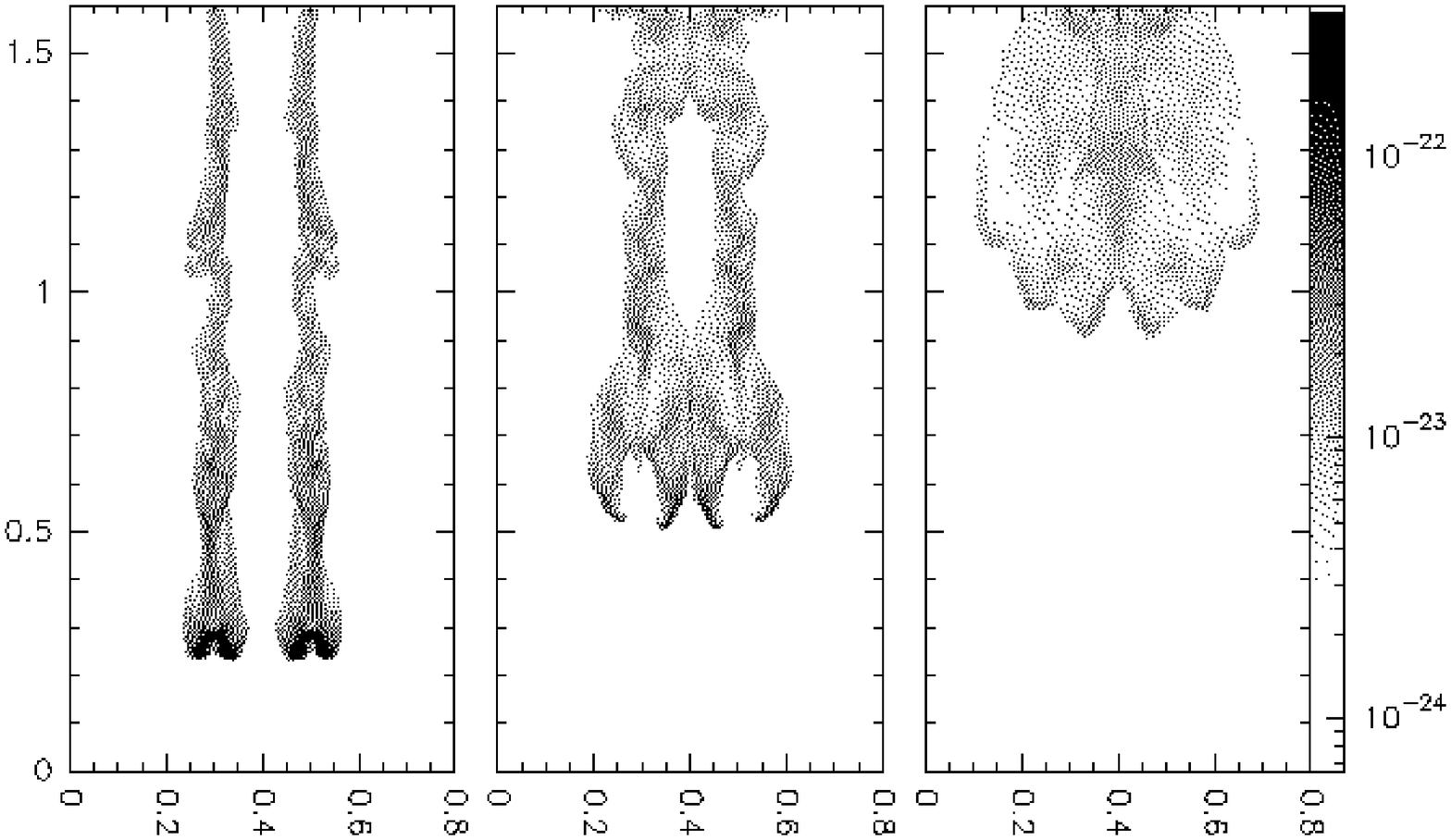}
\end{tabular}
\begin{tabular}{cc}
\epsfxsize=8cm
\epsfbox{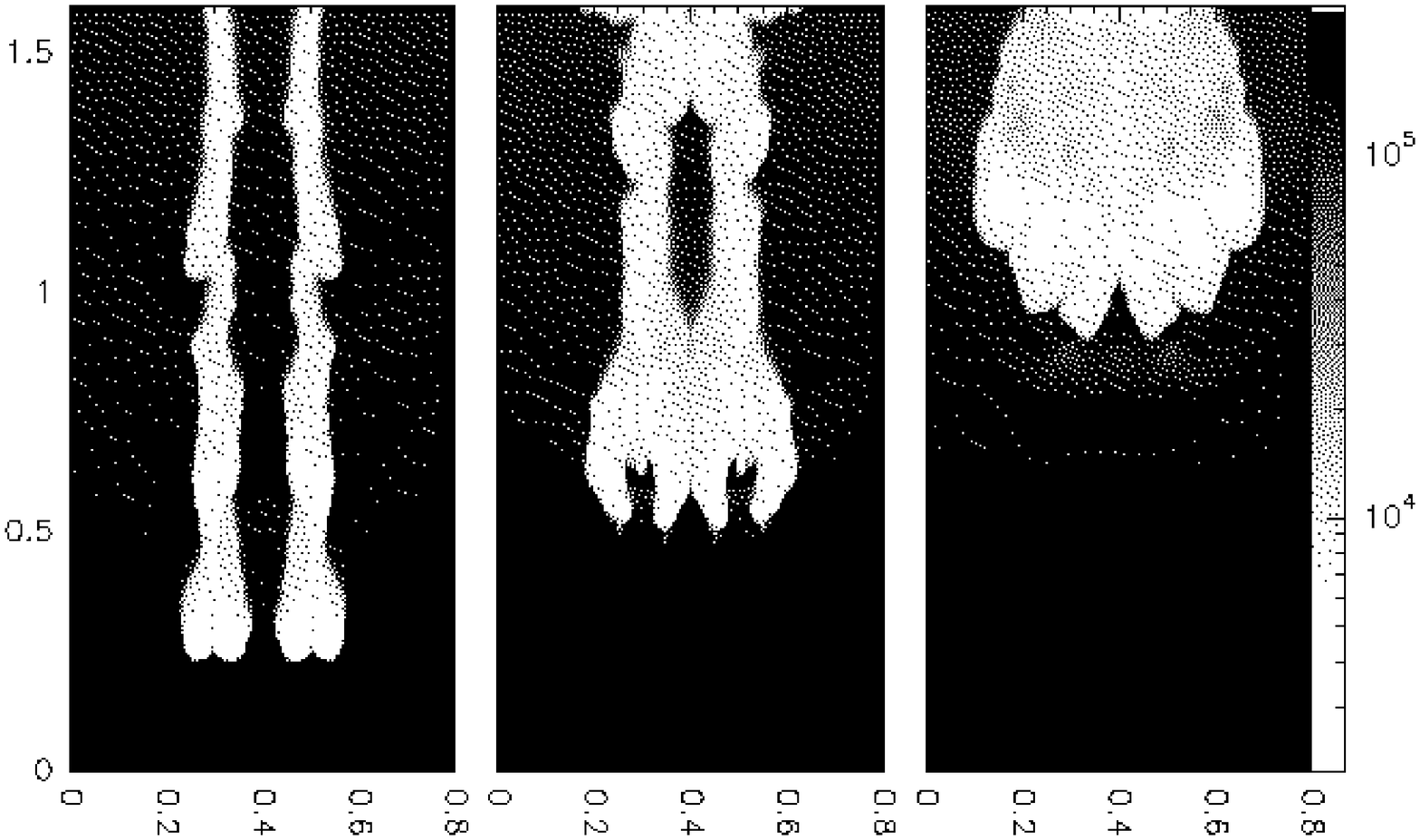}
\end{tabular}
\caption{Model SR2. The same as in Figure 4, except that here it is
considered the radiative cooling
interaction between a SSSF and two clouds.
Time steps are $t$ = 22 $t_{SC}$ (a),
$t$ = 36 $t_{SC}$ (b), and $t$ = 46 $t_{SC}$ (c).}
\end{center}
\end{figure}
\noindent
\begin{figure}
\begin{center}
\begin{tabular}{cc}
\epsfxsize=8cm
\epsfbox{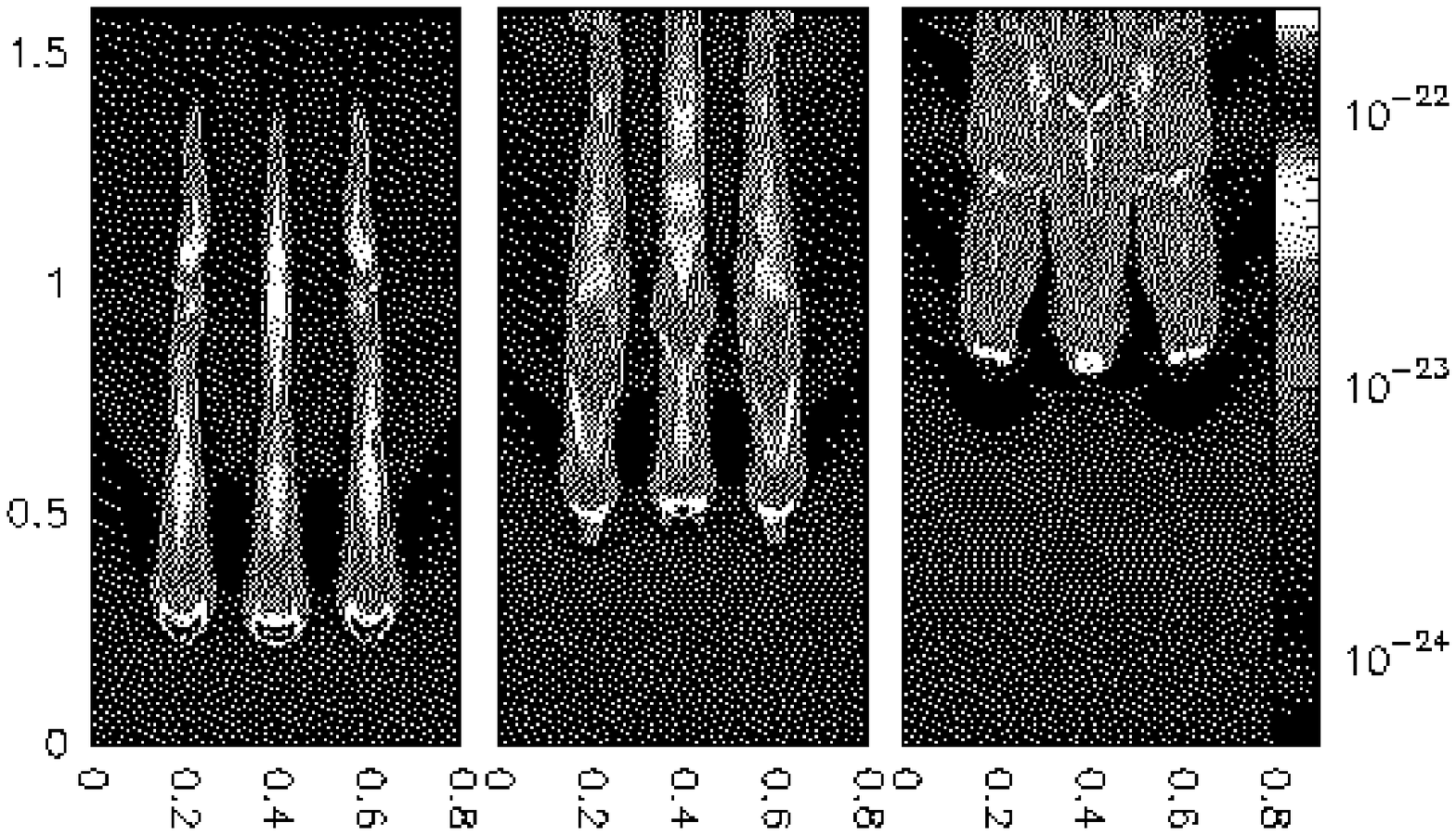}
\end{tabular}
\begin{tabular}{cc}
\epsfxsize=8cm
\epsfbox{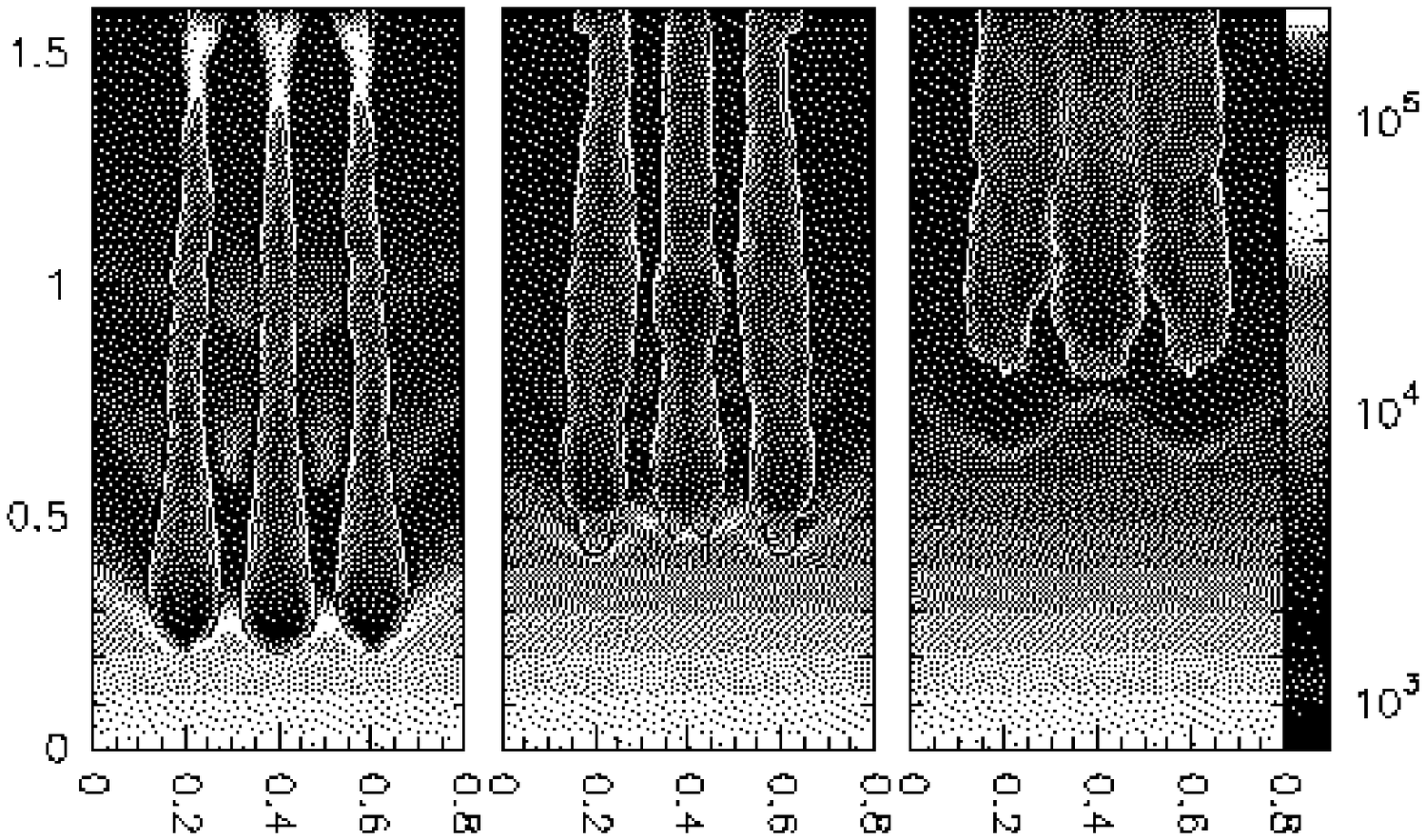}
\end{tabular}
\caption{Model SR3. The same as in Figure 4, except that here we
depict the radiative cooling interaction between an SSSF and
three clouds. Time steps are $t$ = 20 $t_{SC}$ (a), $t$ = 34
$t_{SC}$ (b) and $t$ = 44 $t_{SC}$ (c). In the right panel
we show the Y-Z midplane density and temperature distribution.}
\end{center}
\end{figure}
\noindent
Unlike the results of the non-radiative model SA1,
after a radiative cooling interaction a cloud is almost hentirely
converted to a long cold, dense filament that has suffered less
mass loss to the diffuse ambient gas. As in the non-radiative case, 
the cloud is essentially compressed and
heated by the internal forward shock developed during the impact.
This increases the cloud temperature to about $10^4$ K. While the
internal shock front compresses the cloud, the external bow shock
wave drags the surface cloud material sideways making it lose
mass. Therefore, during the time of the interaction, two important
effects appear: a continuous mass loss due to the drag of material
from the cloud outer parts by the external shock front, and an
internal compression and re-expansion that modifies the structure
of the cloud core.

In the model SR1, at the last stage of the simulation, the cloud
core acquires a normalized velocity ($v_{c,N}$, eq. 5) of 0.34
(that is, 34 \% of the SSSF velocity). Similarly in both
models SR2 and SR3,  the normalized velocities after 50 $t_{SC}$
is 0.4. Their average
velocities over the simulation are
17.3 km s$^{-1}$, 20 km s$^{-1}$ and 19.5 km s$^{-1}$, for the
models SR1, SR2 and SR3, respectively, which are in good
agreement, for instance, with cloud velocities inferred from the
observations of normal galaxies (see, e.g., Boyce \& Cohen, 1994).
The evolution of the clouds velocity, shown in Figure 7, reveals
that they tend to accelerate slowly at the beginning of the interaction and
faster after $\sim$ 26 $t_{SC}$, when the core of the cloud has
lost mass and is thus accelerated with higher efficiency.

The total amount of mass lost by the clouds to the diffuse ambient
medium is
indicated by the normalized mass loss parameter $M_{l,n}$, defined
in eq. 6.
When only one cloud is interacting with the SSSF (SR1), the cloud
core is almost completely destroyed at a time $t = 50 t_{SC}$.
At $t=20 t_{SC}$, we obtain $M_{l,n}$ = 0.1 and at
$t=34 t_{SC}$, we obtain $M_{l,n}$ = 0.6. At $t=50 t_{SC}$, from the
analysis of the density and the temperature profiles, we find that essentially
all the gas that was in the core of the cloud has been ablated by the SSSF.

Klein, McKee \& Colella (1994) find that the
maximum time for cloud destruction is $t_{d,max} \simeq 3.5
(r_c q^{0.5} v_{sh})$.
The initial conditions of the models above would imply
$t_{d,max} \sim 1.8 \times 10^4$ yr, that is $\sim 19 t_{SC}$,
but we find that the total
mass lost by the cloud in the radiative cooling model SR1 is much
less at that time. This indicates a much less efficient cloud
destruction
as a result of the radiative cooling interaction with the SSSF.

The same occurs in the simulations SR2 and SR3.
In the first case, we find $M_{l,n}$ $\sim$ 0.2 at $t=23
t_{SC}$ and $M_{l,n}$ $\sim$ 0.7 at $t=36 t_{SC}$.
In the second case, we find $M_{l,n}$ $\sim$ 0.1 at $t=20 t_{SC}$ and
$M_{l,n}$ $\sim$ 0.8 at $t=34 t_{SC}$.

At the final stage of the three simulations, the cloud core remnants
have become elongated filaments which are colder and denser than
the ISM, so that a complete
mixing of the cloud material with the ambient gas is not observed.
This point is crucial.
Due to the radiative cooling even the ablated gas from the clouds
does not mix completely with the ISM, but instead remains slightly
colder (T $\sim$ 5000 K) and much denser ($\rho \sim 5 \times 10^{-23}$
cm$^{-3}$) than the diffuse gas. This means that the re-expansion
phase is less efficient here than in the adiabatic case, and 
the mixing between ISM and cloud gas is prevented or postponed.

\begin{figure}
\epsfxsize=8cm
\epsfbox{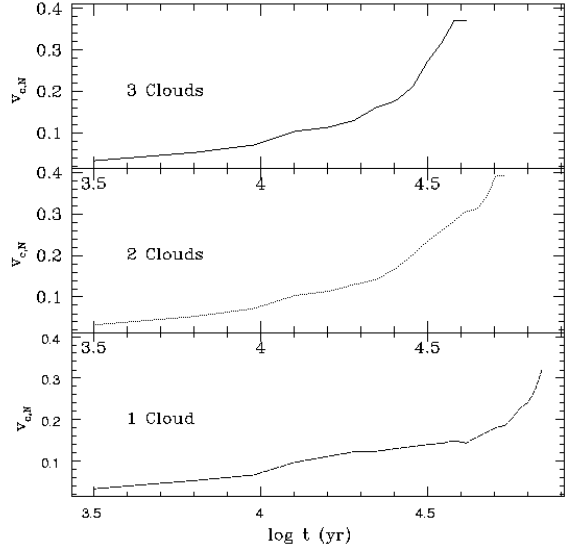}
\caption{Velocity evolution of the clouds in the model SR1 (bottom),
SR2 (center) and
SR3 (top). The velocity is normalized to the SSSF velocity ($v_{sh}=104$ km
s$^{-1}$) and the time is expressed in logarithmic scale.}
\end{figure}
\noindent We conclude, therefore, that the interaction between the
radiative cooling  SSSF and the clouds is unable to destroy them
completely. We also find that the number of elongated structures
with a density larger than $1/e$ of the initial cloud density is
always the same in the three models above (Figs. 4, 5 and 6)
(i.e., $f_{f,t}=$ 1 in eq. 7). This means that the cloud core
numbers remain unaltered while the outer gas is dragged by the
passage of the SSSF.
%\begin{figure}
%\epsfxsize=8cm
%\epsfbox{figure/frag1N.ps}
%\caption{Fragmentation factor for the models SR1 (bottom),
%SR2 (center), and SR3 (top).
%The time is in logarithmic scale.}
%\end{figure}
%\noindent

\subsection{Set 2}

In this section we have run models similar to the previous ones,
except that now we have employed the YGUAZUb code, where, besides
the radiative cooling, we have also considered the presence of a
UV flux of photons. In these simulations, before injecting the
SSSF in the computational domain, we have allowed the cloud to be
photoionized by the UV flux for 6 $\times 10^3$ yr.
Figure 8 shows a photo-evaporating cloud embedded in a bath of UV
photons. The UV flux has been injected from the left, from a star
at a distance of 1 pc (that does not lie within the computational
domain), with an effective temperature of 60.000 K and an UV flux
$F_{49}$ = 0.2, where $F_{49}$ is the flux in units of $10^{49}$
photons/s.

\begin{figure}
\centering
\epsfxsize=8cm
\epsfbox{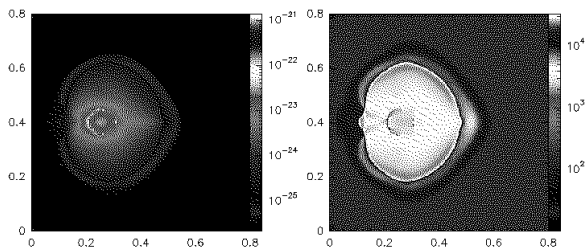}
\caption{Model
SRP1 before the impact with a SSSF. Maps of the
midplane density (left panel) and temperature (right panel)
distribution (in log scale) of a photo-evaporating cloud with $m_c
\sim 0.001 M_{\odot}$, $r_c$=0.05 pc and $\rho_c$ = 1.075 $\times 10^{-22}$
g cm$^{-3}$.
The UV flux is injected from the left of the panel. It has a normalized
value $F_{49}$ = 0.2.}
\end{figure}

%According to the study of Bertoldi and McKee (1989, 1990),
%a cloud
%with the initial conditions of models SRP1 and SRP2 should
%photo-evaporate in $\sim 10^3$ yr, but this does not actually
%occur. In fact,

As we see in Figure 9, the cloud photoionizes
almost completely very rapidly, attaining a temperature in the
core of $\sim 10^4$ K. As a consequence, the cloud becomes
transparent to the UV photon flux, and the photo-evaporation
ceases. At this stage, the only physical phenomenon responsible
for the evolution of the cloud is the free thermal  expansion. When
an SSSF impacts a  cloud in such conditions, a forward shock
front with a velocity lower than the SSSF but with a higher
density develops. This shock front may be seen in Fig. 9 with a
mass density $\sim$ 6.4 $\times 10^{-23}$ g cm$^{-3}$,
or 75 times the number density of the SSSF injected at t = 6000 yr.

We note that this is substantially different from the case
where an SSSF interacts with a neutral $photoionizing$ cloud. In
the present analysis the cloud is already fully photoionized when
it is impacted by the SSSF, and the  UV photons flux that
continues to be injected keeps the temperature of the
denser structures high enough, thus favoring a faster re-expansion
that sweeps away density and temperature fluctuations.
For a detailed study of an interaction between a wind and neutral
photoevaporating clouds covering an extensive parameter space, we
refer to the recent work by Raga, Steffen and Gonzalez (2005).

\begin{figure}
\center
\epsfxsize=8cm
\epsfbox{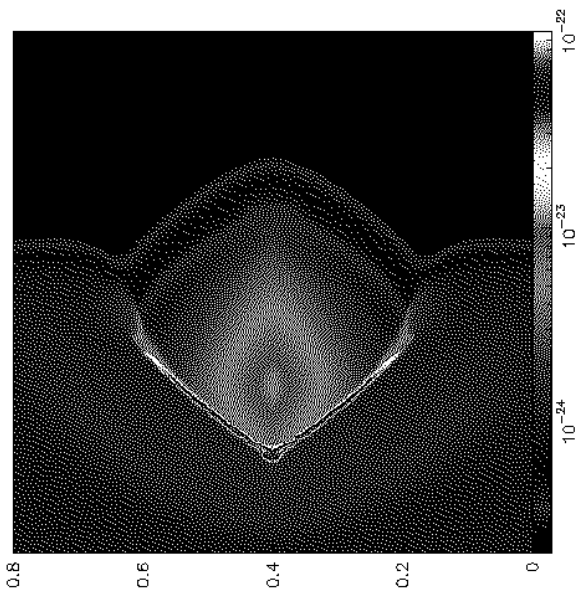}
\caption{Density
profile (in log scale) of model SRP1 at a time $t$ = 3.3 $t_{SC}$.
The initial conditions are the same as in Figure 4, except that
now the cloud is also under the action of a continuous
UV photon flux that comes from the bottom of the box,
with $F_{49}$ = 0.2. The forward shock front resulting from the
interaction between the SSSF and the photo-evaporating gas cloud
has a mass density of $\sim 6.4 \times 10^{-23}$ g cm$^{-3}$.}
\end{figure}

\begin{figure}
\begin{center}
\begin{tabular}{cc}
\epsfxsize=8cm
\epsfbox{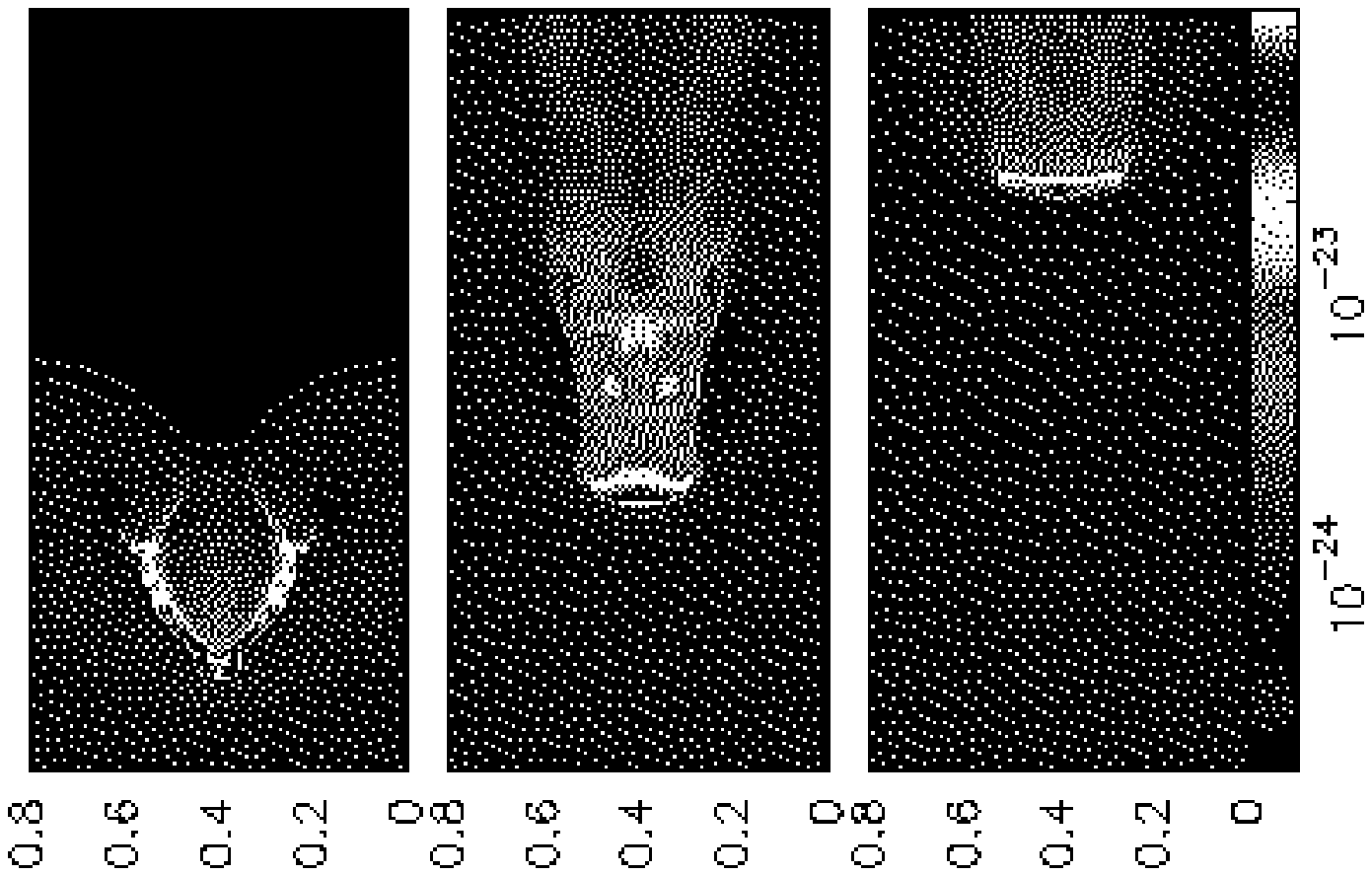}
\end{tabular}
\begin{tabular}{cc}
\epsfxsize=8cm
\epsfbox{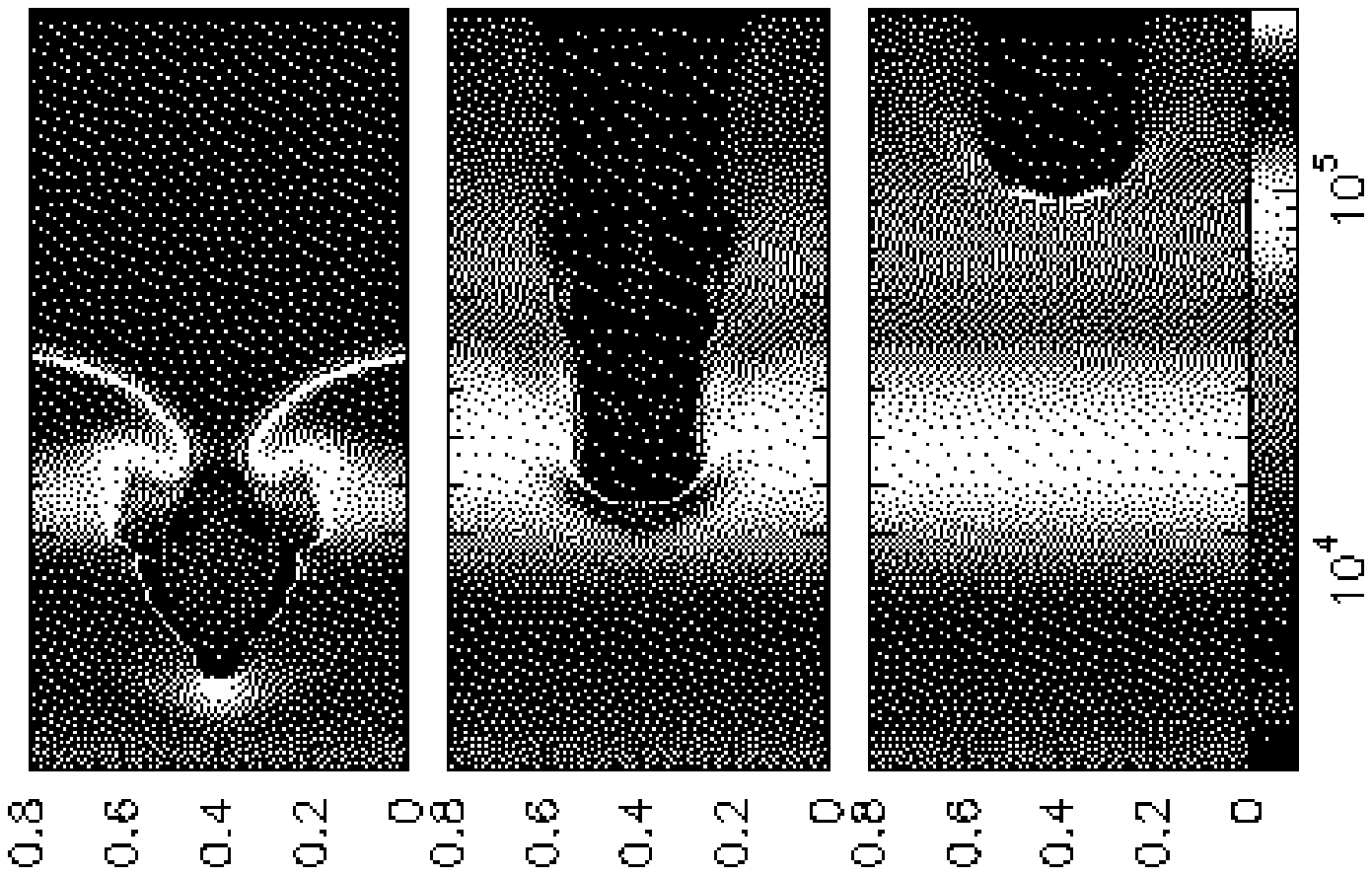}
\end{tabular}
\caption{Later evolution of model SRP1. Color-scale map of the
midplane density (top panel) and temperature (bottom panel)
distributions (in log scale) for the radiative cooling interaction
of an SSSF with a photo-evaporating cloud. The physical conditions
are the same as in Figure 9. Time steps are $t$ = 6.8 $t_{SC}$
(a), $t$ = 20.3 $t_{SC}$ (b), and $t$ = 33.7 $t_{SC}$ (c).}
\end{center}
\end{figure}

\begin{figure}
\begin{center}
\begin{tabular}{cc}
\epsfxsize=8cm
\epsfbox{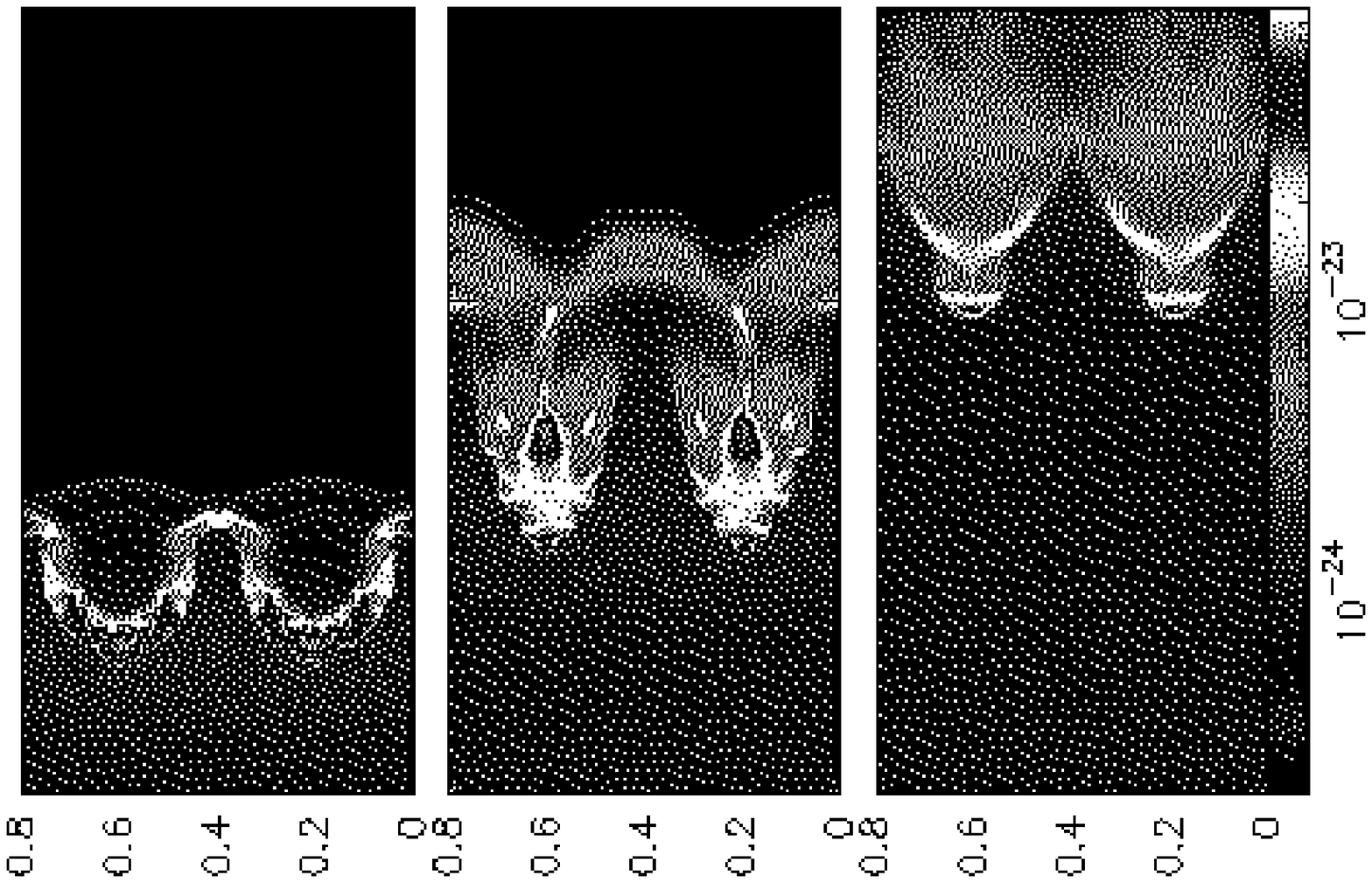}
\end{tabular}
\begin{tabular}{cc}
\epsfxsize=8cm
\epsfbox{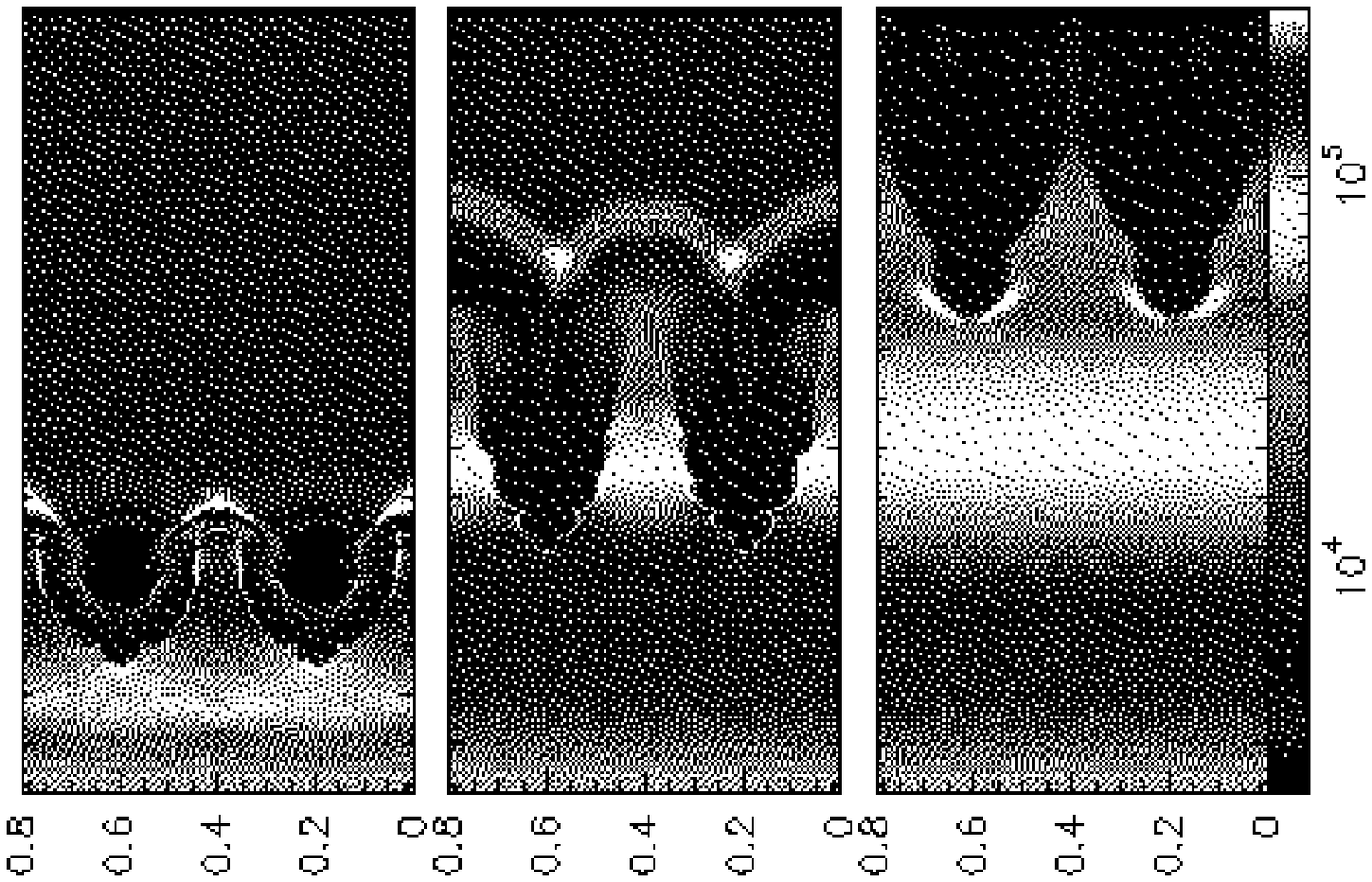}
\end{tabular}
\caption{Model SRP2. The same as in Figure 9, except that now the
radiative cooling interaction between an SSSF and a system of two
photo-evaporating clouds is considered. Time steps are $t$ = 6.8
$t_{SC}$ (a), $t$ = 14 $t_{SC}$ (b), and $t$ = 23 $t_{SC}$
(c).}
\end{center}
\end{figure}

The density and the temperature evolution of models SRP1 and
SRP2, with one and two photoevaporated clouds, respectively, are
plotted in Figures 10 and 11. In these cases, the interaction
between the forward shock front and the clouds is stronger than in
the cases without photo-evaporation, as the effective density
contrast between the cloud and the SSSF is only $\sim$ 1.7, or 80
times lower than in the models of Set 1 without photo-evaporation.
As a consequence, the clouds are pushed and expand with higher
efficiency. The normalized velocity at a time of 33.7 $t_{SC}$ is
0.5 for the run SRP1 and 0.57 for the run SRP2. These values are
higher than those obtained in the runs of Set 1, and are
reached in a shorter time. In Figure 12, the evolution of the
normalized velocity is shown for both cases. Due to the faster
acceleration, the resulting average velocity is 36.7 km s$^{-1}$
for the cloud of model SRP1 and 44.5 km s$^{-1}$ for the clouds
of model SRP2.

\begin{figure}
\epsfxsize=8cm
\epsfbox{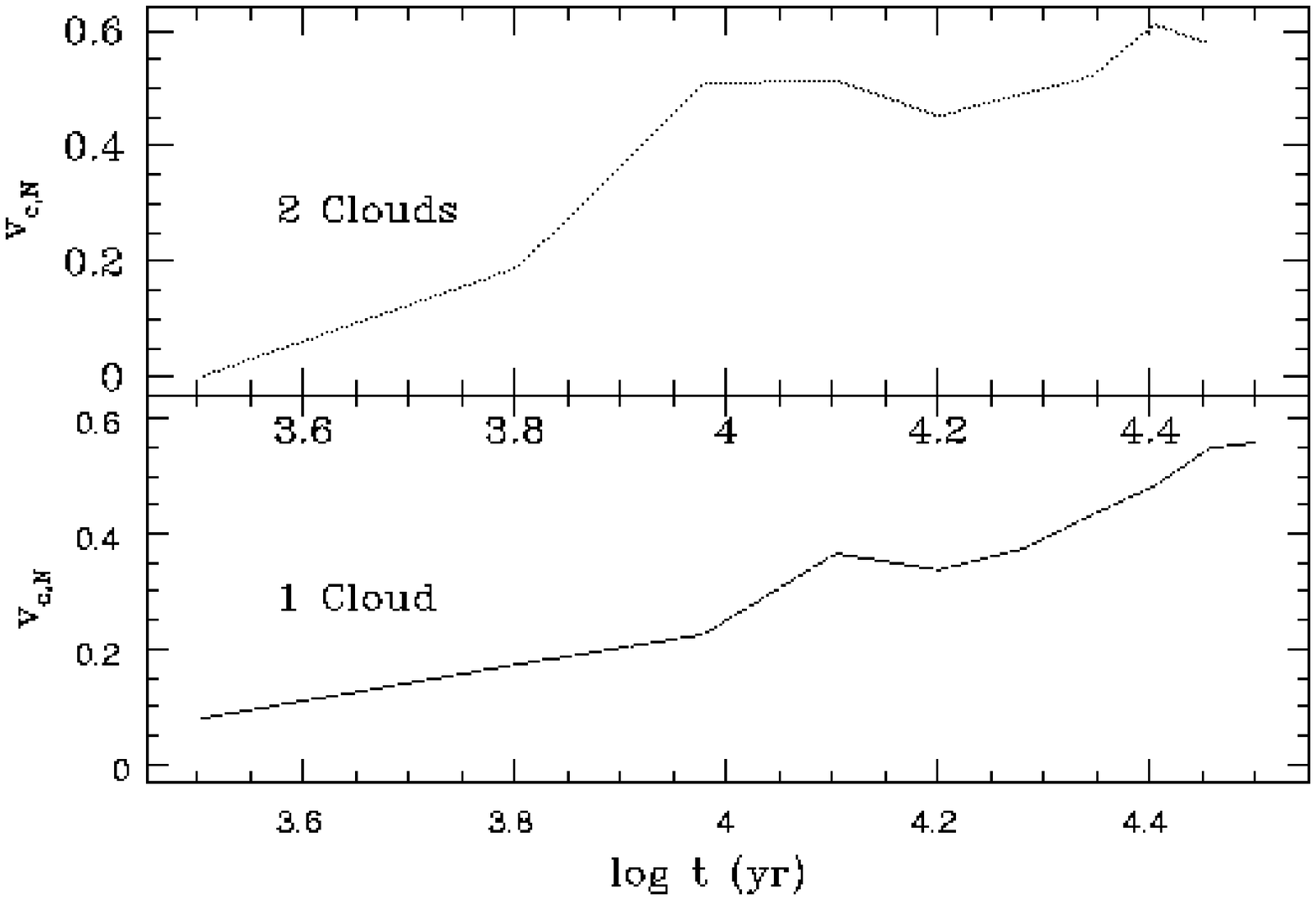}
\caption{Velocity
evolution of the photoionized clouds of the models SRP1 (bottom),
and SRP2 (top). The velocity is normalized to the SSSF velocity,
and the time is expressed in logarithmic scale.}
\end{figure}

When the SSSF wind is injected, the maximum density of the
photo-evaporating clouds is $\sim 2.15 \times 10^{-23}$ g cm$^{-3}$,
that is only 1/5 of the initial cloud density.
Taking this value, $\rho_{c,ph}$, as
the initial density of the photionized cloud at the time of the
interaction with the SSSF, we find that the total mass of the core
with $\rho \ge \rho_{c,ph}/e$ is only 35\% of the initial cloud mass.
The rest of the gas spreads in the ISM, with an average mass
density of $\sim 2.15 \times 10^{-24}$ g cm$^{-3}$ and a volume which is 
sixty times
larger than the initial volume. Later, as it interacts with the
SSSF, the shocked cloud gas is pushed away and expands more
efficiently, but even in this case the mixing with the ISM is not
efficient and no significant fragmentation is observed (Figures 10
and 11). The normalized fragmentation factor (as defined in eq. 7)
is always 1 for both  models SRP1 and SRP2, which is the same
found for the models of the set 1.

%\begin{figure}
%\epsfxsize=8cm
%\epsfbox{figure/fragf.ps}
%\caption{Normalized
%fragmentation factor for the models SRP1 (bottom), and SRP2 (top).
%The time is expressed in logarithmic scale.}
%\end{figure}
%\noindent

\subsection{Set 3}

In the third set of simulations, instead of a steady state shock
front (SSSF), we have considered the interaction of a SNR shell
with one (models SNS1 and SNSP1) and two clouds (models SNS2 and
SNSP2). The simulations without a UV photon flux (SNS1 and SNS2)
were run with the YGUAZUa version of the code, and the simulations
with a continuous flux of UV photons (SNSP1 and SNSP2) were run
with the YGUAZUb code. The results are shown in Figures 13 to 16.

\begin{figure}
\begin{center}
\begin{tabular}{cc}
\epsfxsize=8cm
\epsfbox{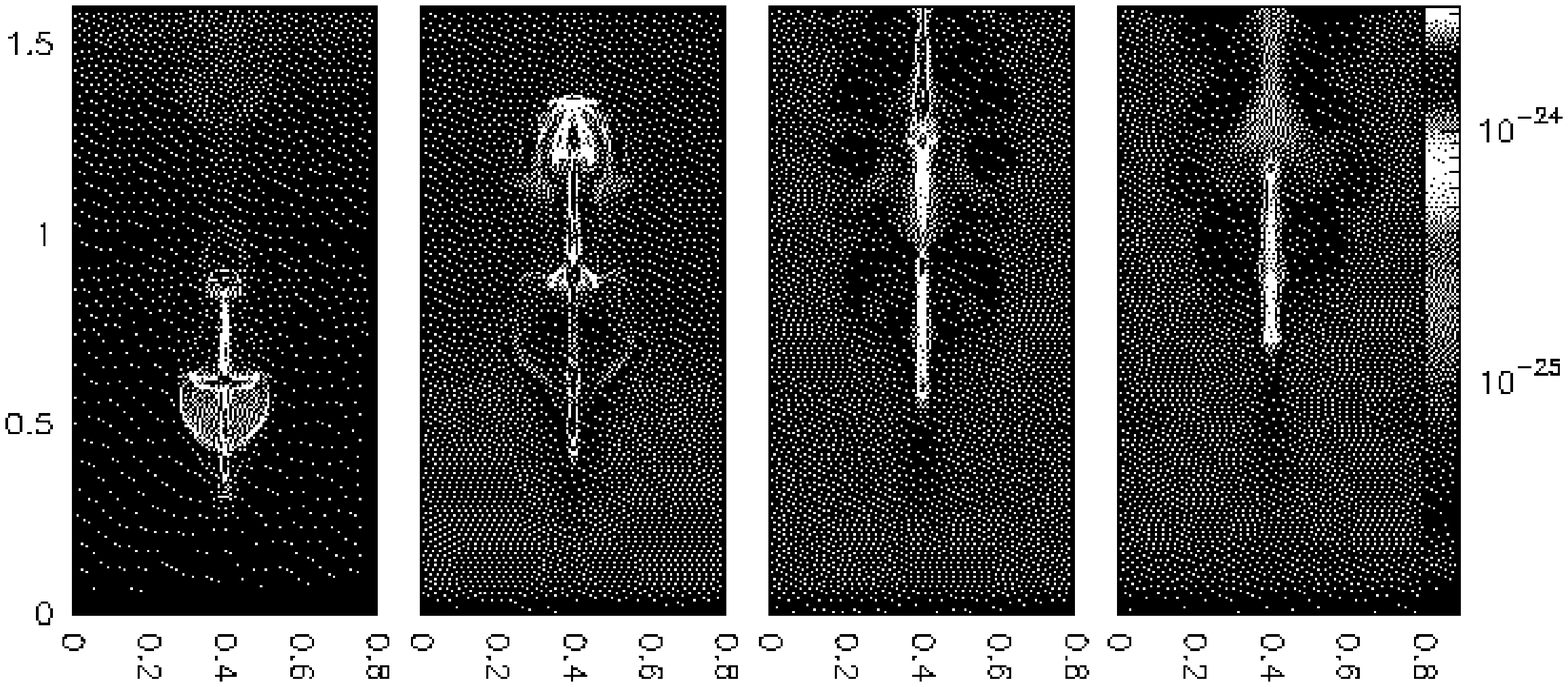}
\end{tabular}
\begin{tabular}{cc}
\epsfxsize=8cm
\epsfbox{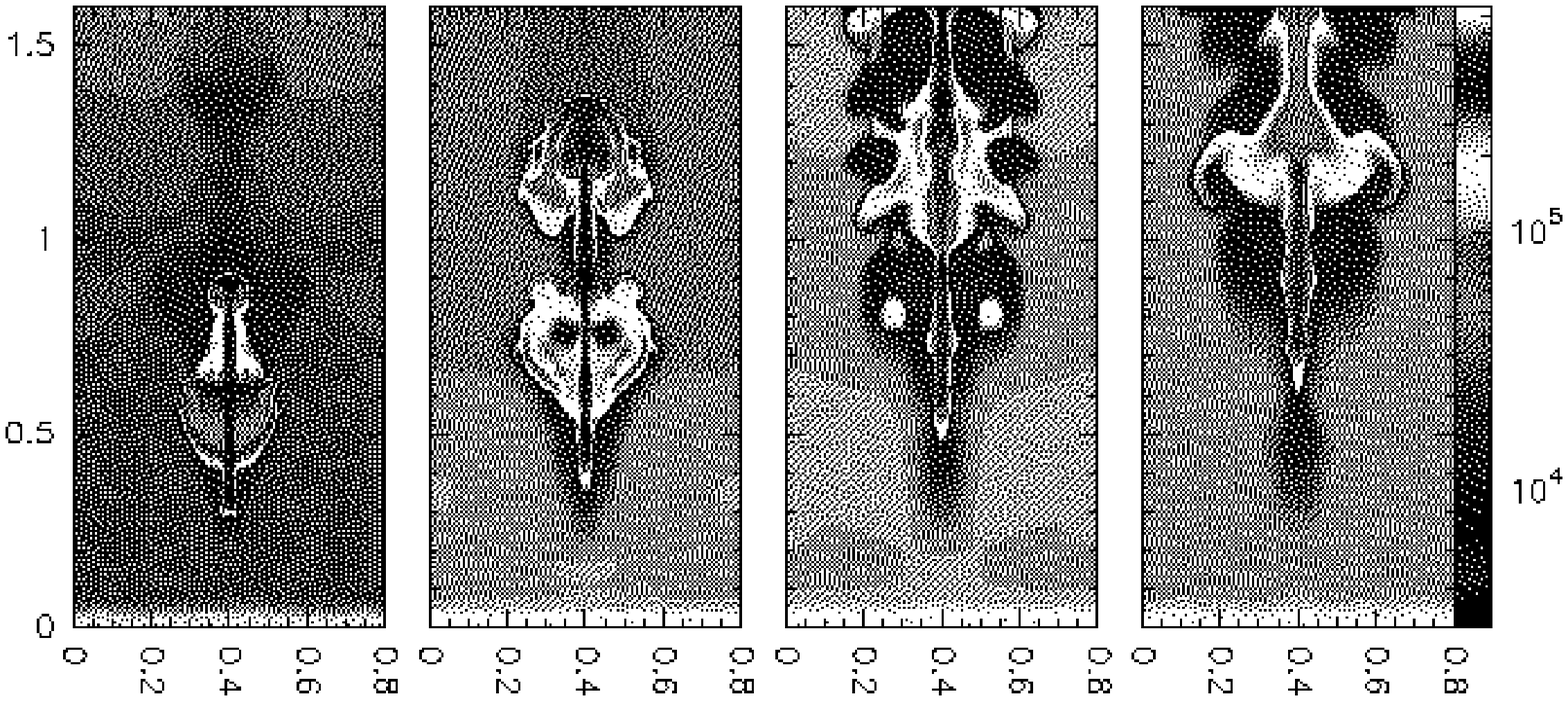}
\end{tabular}
\caption{Model SNS1. Maps of the midplane density (top
panel) and temperature (bottom panel) distributions (in log scale)
for the interaction of a SNR with one cloud. The cloud has a
mass density $\rho_c$ = 2.15 $\times 10^{-24}$ g cm$^{-3}$,
a temperature T = 100 K and a
radius $r_c$ = 0.05 pc, and is embedded in an ISM which has a
mass density $\rho_a$ = 2.15 $\times 10^{-26}$ g cm$^{-3}$
and a temperature T = $10^4$
K. The SNR is injected from the bottom of the box with a velocity
of 250 km s$^{-1}$, a mass density $\rho_{sh}$ = 8.6 $\times 10^{-26}$ g
cm$^{-3}$
and a temperature T= {\bf $8.2 \times 10^5$} K. Time steps are $t$ = 48
$t_{SC}$ (a), $t$ = 81 $t_{SC}$ (b), $t$ = 121 $t_{SC}$ (c), and
$t$ = 162 $t_{SC}$ (d).}
\end{center}
\end{figure}
\noindent
\begin{figure}
\begin{center}
\begin{tabular}{cc}
\epsfxsize=8cm
\epsfbox{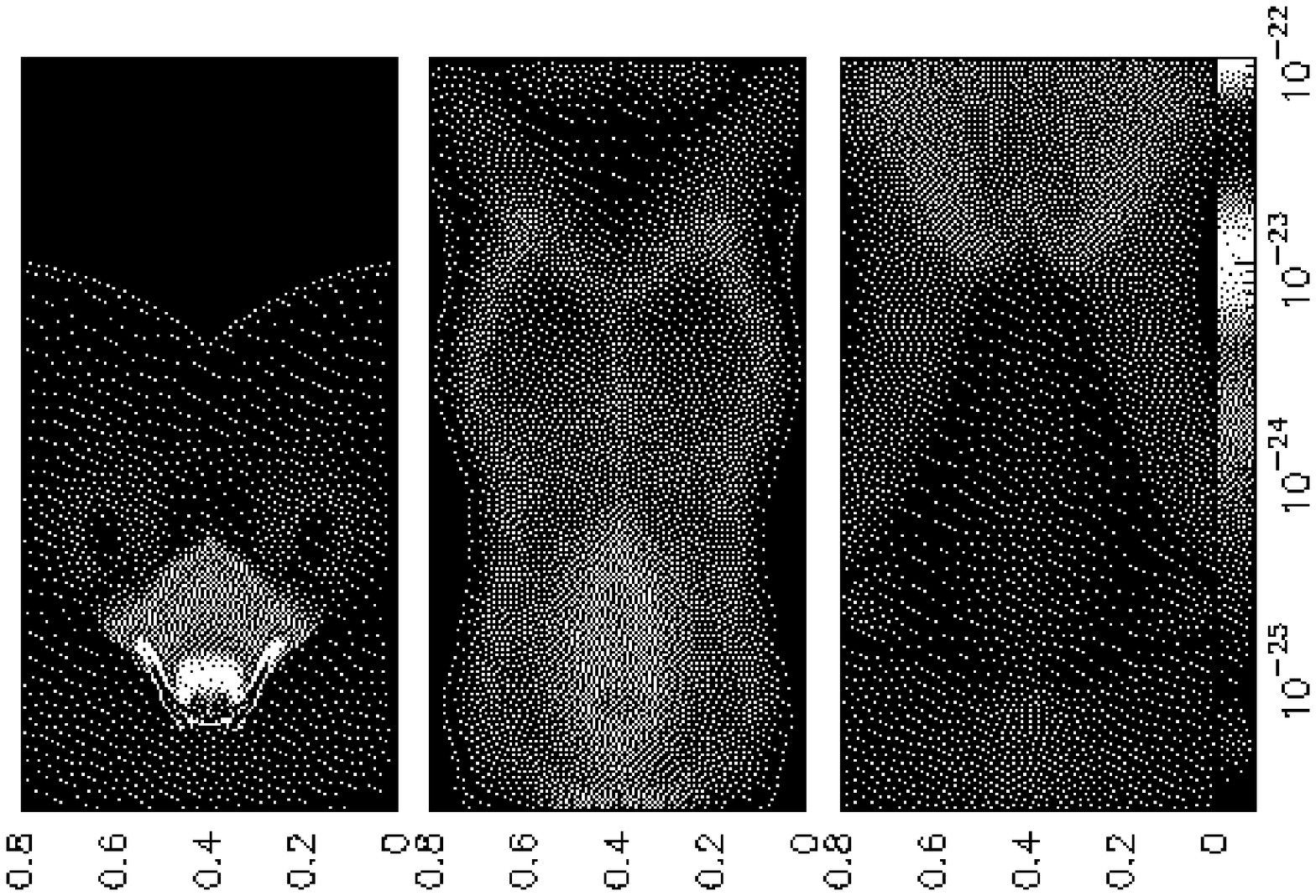}
\end{tabular}
\begin{tabular}{cc}
\epsfxsize=8cm
\epsfbox{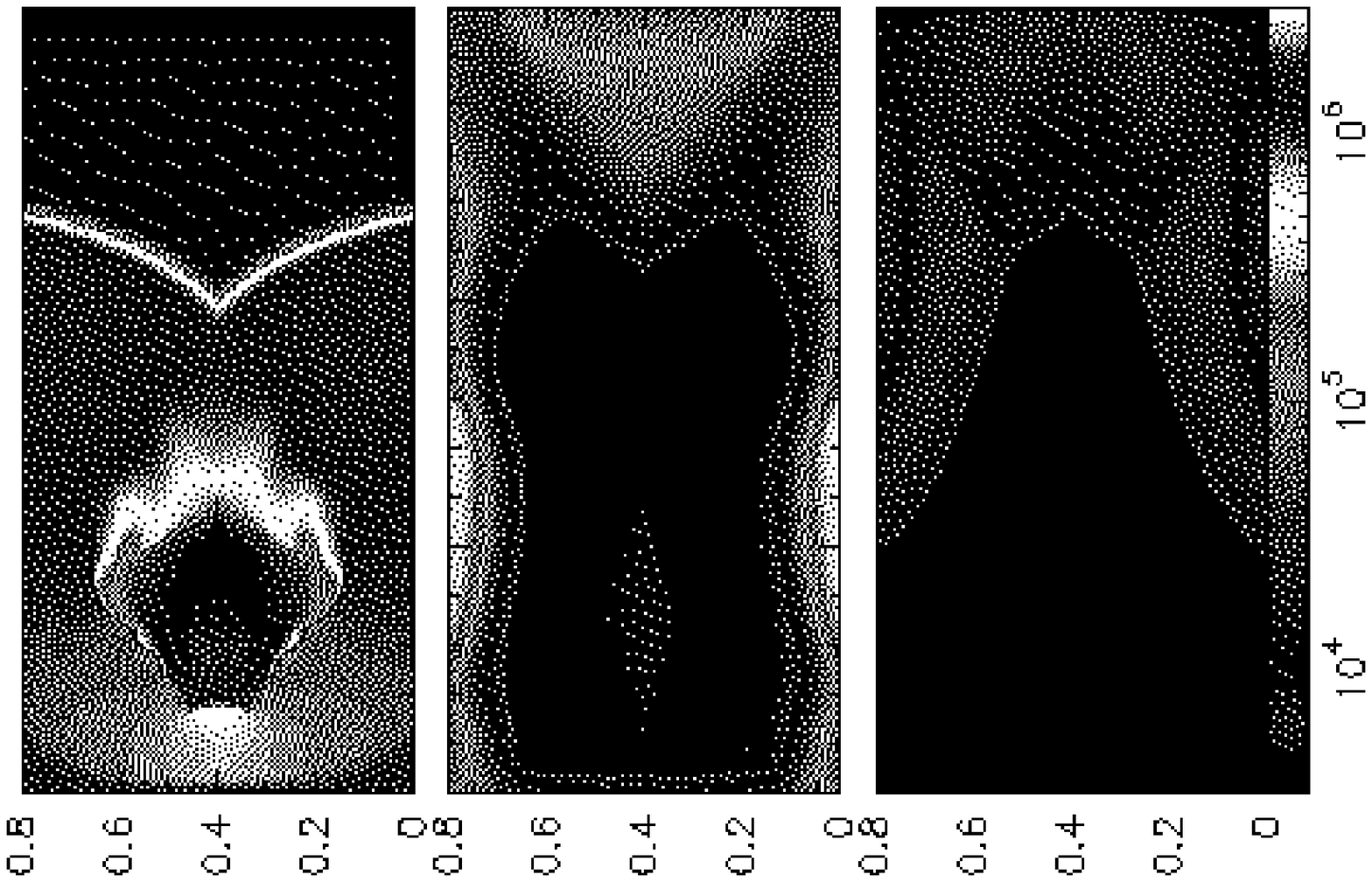}
\end{tabular}
\caption{Model SNSP1. Color-scale map of the midplane density (top
panel) and temperature (bottom panel) distributions (in log scale)
for the interaction of an SNR with one photo-evaporating cloud. The
cloud has a mass density $\rho_c$ = 1.075 $\times 10^{-22}$ g
cm$^{-3}$, a temperature T =
100 K and a radius $r_c$ = 0.05 pc and is embedded in an ISM
with a mass density $\rho_a$ = 2.15 $\times 10^{-26}$ g
cm$^{-3}$ and a temperature
T = $10^4$ K. The SNR is injected from the bottom of the box with
a velocity of 250 km s$^{-1}$, a mass density $\rho_{sh}$ = 8.6
$\times 10^{-26}$ g
cm$^{-3}$ and a temperature T= {\bf $8.2 \times 10^5$} K. Time
steps are $t$ = 8.1 $\times 10^3$ yr (a), $t$ = 48.7 $t_{SC}$ (b),
and $t$ = 129 $t_{SC}$ (c).}
\end{center}
\end{figure}
\noindent
\begin{figure}
\begin{center}
\begin{tabular}{cc}
\epsfxsize=8cm
\epsfbox{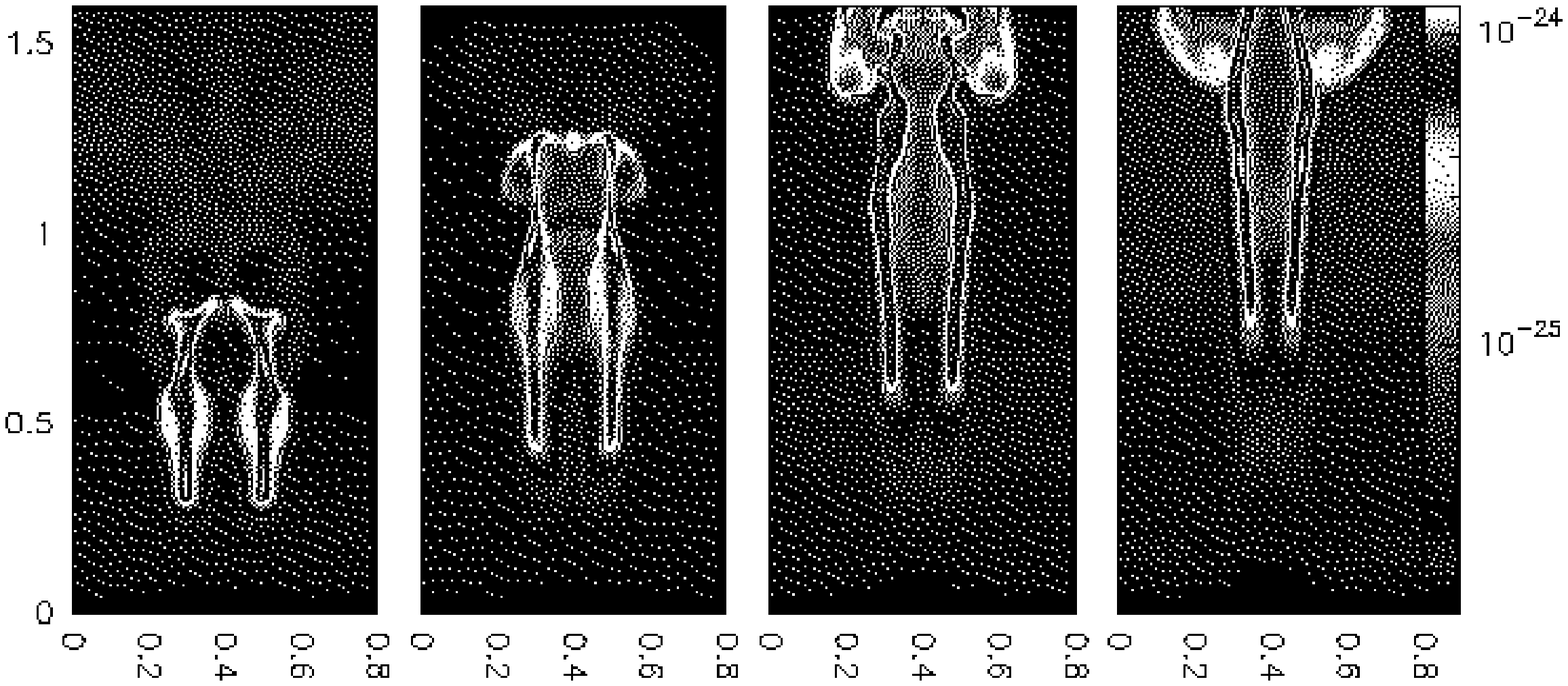}
\end{tabular}
\begin{tabular}{cc}
\epsfxsize=8cm
\epsfbox{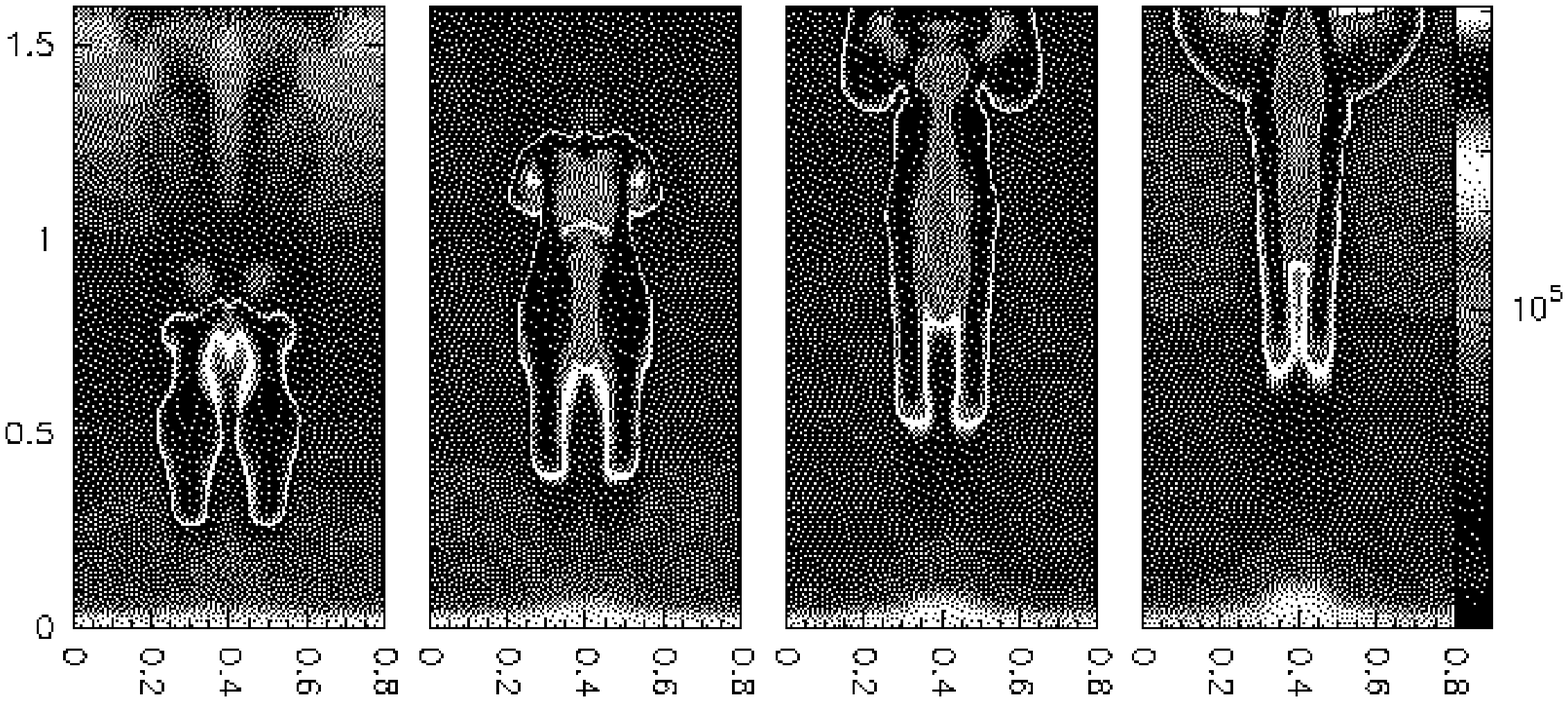}
\end{tabular}
\caption{Model SNS2. The same as in Figure 15, except that now the
radiative cooling interaction between an SNR and two clouds is
considered. Time steps are $t$ = 48 $t_{SC}$ (a), $t$ = 89
$t_{SC}$ (b), $t$ = 129 $t_{SC}$ (c) and $t$ = 162 $t_{SC}$ (d).}
\end{center}
\end{figure}
\noindent
\begin{figure}
\begin{center}
\begin{tabular}{cc}
\epsfxsize=8cm
\epsfbox{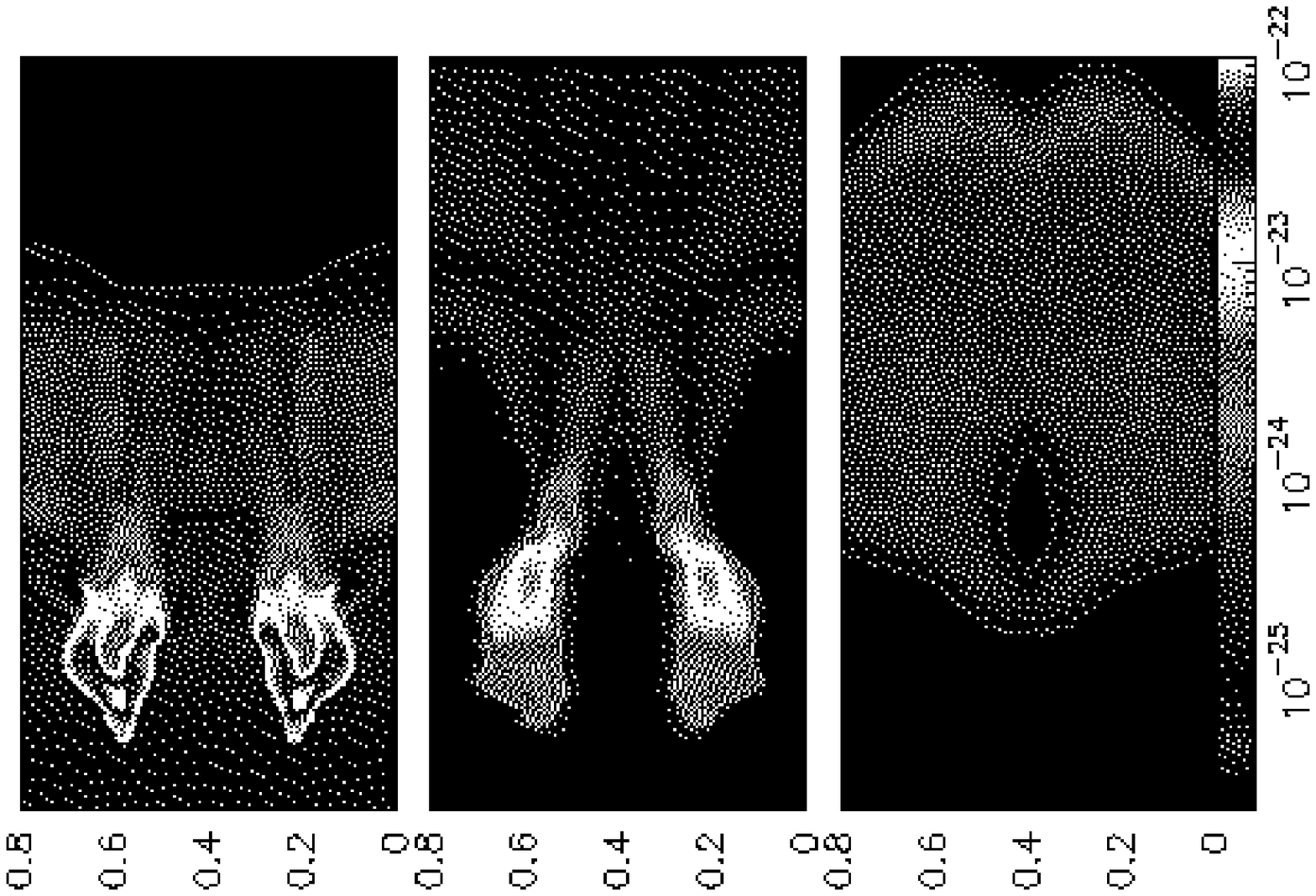}
\end{tabular}
\begin{tabular}{cc}
\epsfxsize=8cm
\epsfbox{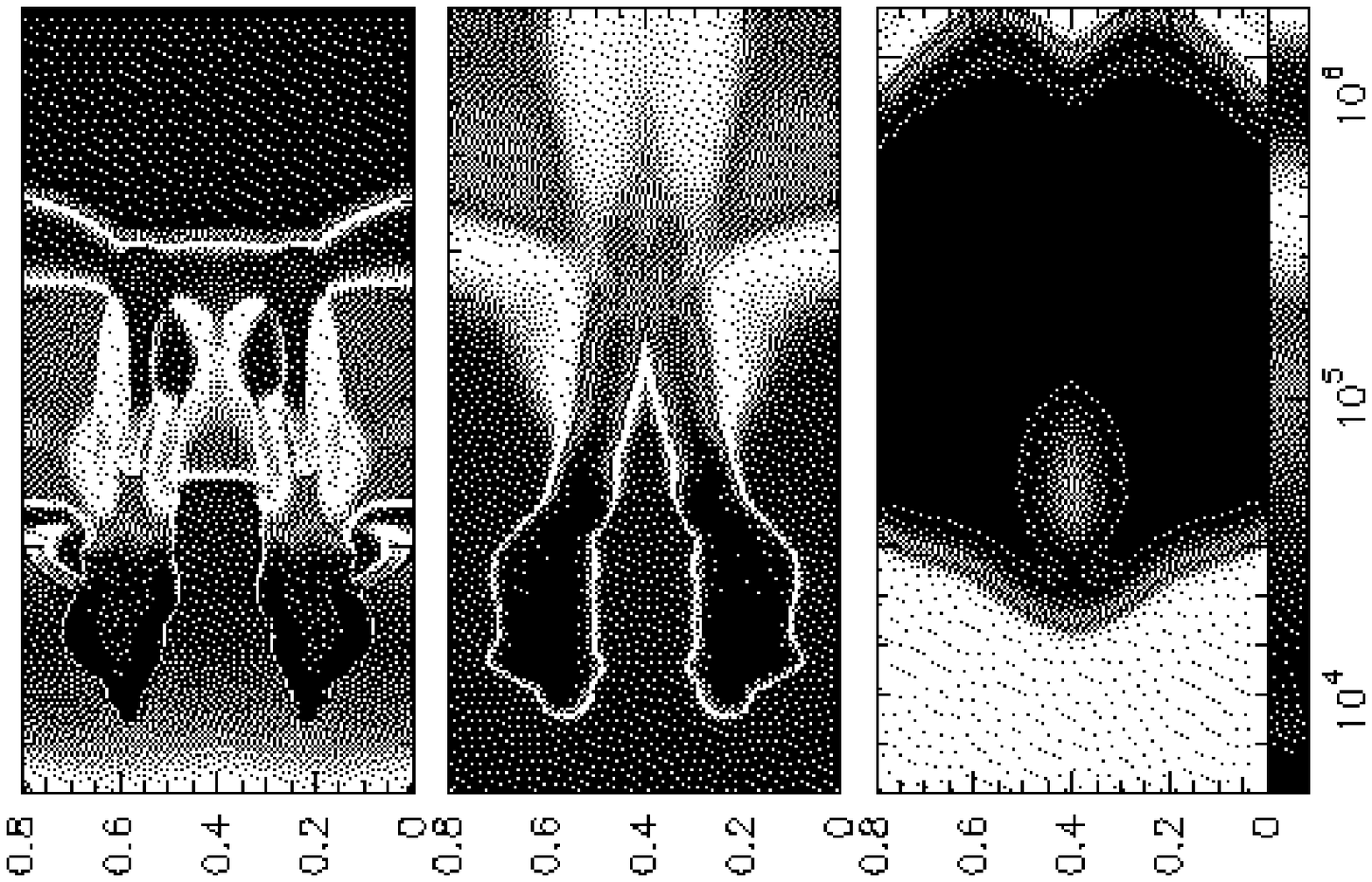}
\end{tabular}
\caption{Model SNSP2. The same as in Figure 16, except that now
the radiative cooling interaction between an SNR and two
photo-evaporating clouds is considered. Time steps are $t$ = 6.5
$\times 10^3$ yr (a), $t$ = 13 $t_{SC}$ (b), $t$ = 27.6 $t_{SC}$
(c) and $t$ = 97.4 $t_{SC}$ (d).}
\end{center}
\end{figure}
\noindent In this set, we have considered initial conditions
appropriate for an SB galaxy environment. Due to the high rate
of SN explosions, the ISM is at low density and high temperature,
and thus, the SNR shells are expected to have much lower
densities than that of the SSSF considered in Sets 1 and 2 (we
assume here an SNR shell with a density ten times smaller). The
small clouds, presumably formed by the fragmentation of shells
formed in an earlier generation of SNe that exploded in the SB
ambient (Melioli and de Gouveia Dal Pino 2004), have a mass density
$\rho_c$ = 2.15 $\times 10^{-24}$ g cm$^{-3}$ in the models without
photo-evaporation, SNS1
and SNS2, and $\rho_c$ = 1.075 $\times 10^{-22}$ g cm$^{-3}$
in the models with
photo-evaporation, SNSP1 and SNSP2. This difference between the
initial densities of the clouds is justified by the fact that the
density contrast between the clouds and the ambient medium is
expected to be higher in photoionized regions (see, e.g., Bertoldi
\& McKee 1990). In the presence of UV photons, the initial
internal pressure of the cloud must be higher than the ISM
pressure ($\sim$ 10 times) and this forces the expansion of the
cloud in the photoionized environment. Thus, for the models
SNS1 and SNS2, we take a density contrast five times lower than in
the cases studied in the Sets 1 and 2, while for the models SNSP1
and SNSP2, we take a density contrast 10 times higher. The shell
is assumed to have a thickness of 1 pc (see, e.g. McCray 1992). 
Aiming at studying more realistic systems involving
SNR-clouds interactions, the parameters we have adopted here
such as that the clouds have dimensions that are not much smaller
than the thickness of the SNR shell, $r_c \sim h_{sh}/
q^{0.5}$ (see definition of a small cloud by Klein, McKee \&
Colella 1994). This justifies in part some of the differences
to the results found in the sets 1 and  2, and also explains
some differences with respect to analytical solutions, as we will
discuss in the next section.

In the cases without photo-evaporation (Figures 13 and 15), the
filamentation of the clouds and gas mixing with the ISM due to the
interaction with the SNR is less efficient than in the previous
sets with an SSSF, in spite of the smaller density contrast between
the cloud and the shell ($q = \rho_c/\rho_{sh}$ = 25 in Figs.
13 and 15) than in the models of the sets 1 and 2 (where $q =
\rho_c/\rho_{sh}$ = 125). This is explained by the fact remarked
above that in these models of Fig. 13 and 15 (Set 3) the thickness
of the shock front is not much larger than the size of the clouds.
In Figures 13 and 15, the gas mixing timescale must be much longer
than the total computation time which is of the order of 5 $\times
10^5$ yr.

As in Set 1, we can also estimate  the radiative cooling
timescales behind the double shock structure that develops at the
contact surface  between the SNR shell and each cloud after the
impact. For a shell velocity $v_{sh} = 250 $ km s$^{-1}$ and a
density contrast $q =$ 25, we find that the forward shock speed
into the cloud is $v_{s,c} \simeq 50$ km s$^{-1}$ and the reverse
shock speed into the SNR shell is $v_{s,SNR} \simeq 200$ km
s$^{-1}$. Thus the radiative cooling time of the shocked gas in
the SNR shell is $t_{c,SNR} \sim 4.1 \times 10^5  \ {\rm yr} \
(v_{s,SNR} / 200 \ {\rm km \ s^{-1}})^{1.12} (\rho_{SNR} / 8.6
\times 10^{-26} \ {\rm g cm^{-3}})^{-1}$, which is much longer
than the cloud crushing time, $t_{crush} \sim t_{SC} \ q^{0.5}
\simeq 1.9 \times 10^3$ yr. This long cooling time is due to the
very low density of the shell material. An even longer radiative
cooling time is found for the shocked ambient material ($\sim 1
\times 10^6$ yr) which is more rarified than the shell. On the
other hand, the radiative cooling time of the shocked material in
the cloud is much smaller, $t_{c,c} \sim 1.8 \times 10^4  \ {\rm
yr} \ (v_{s,c} / 50 \ {\rm km \ s^{-1}})^{-3.58} (\rho_{c} / 2.15
\times 10^{-24} \ {\rm g \ cm^{-3}})^{-1} \simeq 9 t_{crush}
\simeq 47 t_{SC}$. A though longer than the cloud crushing time,
this time scale is short enough to provide the observed radiative
cooling of the shocked cloud material in the models of Figs. 13
and 15 within a few crushing times. We can also estimate the
thickness of the cold shell that develops due to the cooling of
the shocked material around each cloud. For a shock velocity
smaller than 80 km s$^{-1}$ (Hartigan et al. 1987),  $d_{c,c}
\simeq 0.3 r_c \ (v_{s,c} / 50 \ {\rm km \ s^{-1}})^{-4.51}
(\rho_{c} / 2.15 \times 10^{-24} \ {\rm g \ cm^{-3}})^{-1}$, which
is compatible with the thickness of the cold material in the
simulations (see Figs. 13 and 15).

The clouds in Figs. 13 and 15 are less accelerated by the shock
front than in the sets with an SSSF and develop a very elongated
filamentary structure with a temperature of $\sim 5 \times 10^3$
K and a density of $\sim$ 2.15 $\times 10^{-24}$ g cm$^{-3}$
that grows with time.
The normalized velocity of the elongated filaments
are only 0.04 after a time of 129 $t_{SC}$ for both models SNS1 and SNS2. The
evolution of the cloud velocity is shown in Figure 17. For  model
SNS1, after 162 $t_{SC}$ (the total simulated interval), we obtain
a normalized mass loss to the ISM of only $M_{l,n}$ = 0.5, so that
half of the mass of the cloud is still in the filaments. For model
SNS2, we obtain, after 56 $t_{SC}$, $M_{l,n}$ = 0.2, and after 162
$t_{SC}$, $M_{l,n}$ = 0.8. In these
cases no significant fragmentation of the elongated structures is
produced during the simulations. In these cases, the
elongated filaments
have a thickness of $\sim $ 0.03 pc; as we have seen in eq. (9),
the most disruptive Kelvin-Helmholtz (K-H) modes have wavelengths
corresponding to $kr_c \simeq 1$. If we substitute $r_c$ by the
half-thickness of the filament, we obtain that the most
disruptive wavelength is $\lambda = 2\pi/k \sim$ 0.1 pc which
corresponds to 64 grid points. This indicates that the absence of
K-H fragmentation is not due
to unresolved physics of our grid but to the presence of
radiative cooling as discussed earlier.

When a UV photon flux is present, the results change considerably
(Figures 14 and 16). When the SNR shell is injected, the cloud
core has already lost 50\% of its initial mass due to
photo-evaporation, and the forward shock wave that is produced
from the interaction of the SNR with this photo-evaporating cloud
is more efficient in destroying it. The interaction accelerates
the cloud to a normalized velocity of 0.08 of the SNR shell
velocity, and since all the cloud is at a temperature greater than
5000 K, the shocked gas begins to expand and fill in the environment
around the cloud. No fragments are observed, and the cloud cores rapidly
disappear in both Figures 14 and 16. In this situation, a real
mixing between the ISM and the cloud gas takes place very rapidly in the SB
environment. In a time of 129 $t_{SC}$, that is $\sim
5 \times 10^4$ yr, the ambient medium increases its mass density
to a value of 2.15-6.4 $\times 10^{-25}$ g cm$^{-3}$.
This is 10 times greater than the
initial ISM density, while the ISM temperature remains of the
order of $10^3-10^4$ K. This result is consistent with that
obtained in previous work by Melioli \& de Gouveia Dal Pino (2004)
where the timescale for mixing of the gas from the clouds with
the ISM in a SB environment due to photo-evaporation was found to
be much smaller ($\sim 10^4$ yr) than the mixing timescale due to
mass loss from SNR-cloud interactions.

\begin{figure}
\epsfxsize=8cm
\epsfbox{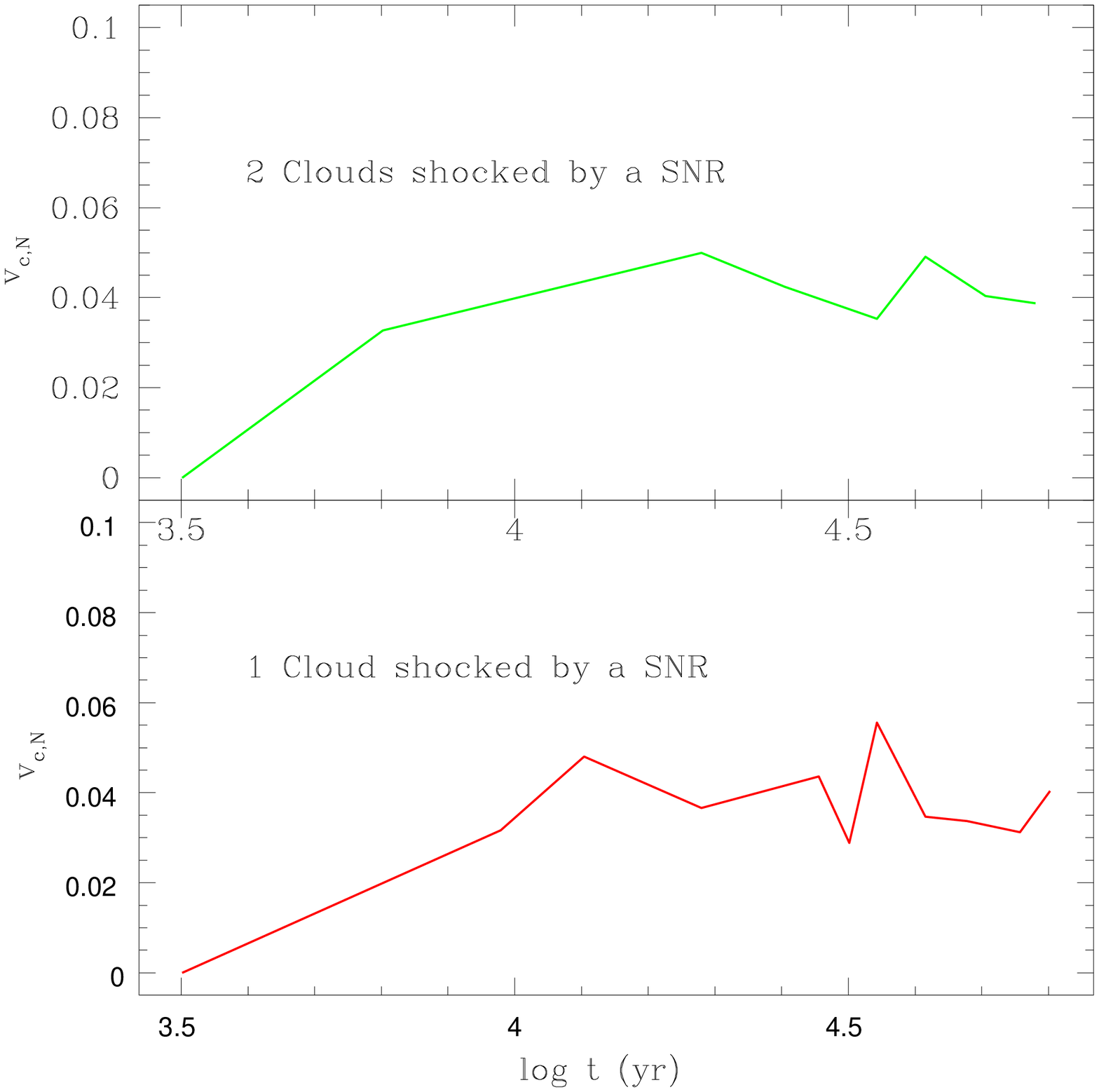}
\caption{Velocity
evolution of the models SNS1 (bottom),  and SNS2 (top).
The
velocity is normalized to the SNR shell velocity, and the time is
expressed in logarithmic scale.}
\end{figure}

\section{Discussion and conclusions}
\subsection{Cloud velocity}

When a cloud is impacted,
its equation of motion has the approximate form:

\begin{equation}
m_c{{dv_c} \over {dt}} \sim v_{sh}^2 \Sigma_c \rho_{sh}
\end{equation}
\noindent where $\Sigma_c$ is the cloud cross section, $m_c$ is
the cloud mass and $\rho_{sh}$ is the density of the shell or
wind that impacts on it. Thus, from eq. 5, the normalized
velocity is:

\begin{equation}
v_{c,N} \sim  {{v_{sh} \Sigma_c \rho_{sh} t} \over {m_c}}
\end{equation}
\noindent where we have assumed that the cloud mass and cross
section and the shock wave density do not change with time. If we
substitute the initial values adopted in our simulations, the
normalized velocity of the clouds is $v_{c,N}\sim 10^{-5} t$ for both sets 1
and 2, and $v_{c,N}\sim 1.5 \times
10^{-4} t$ for set 3, where the time is expressed in years.

This means that after a time of
$\sim 5 \times 10^4$ yr the cloud should have a velocity $\sim$
0.5 times the shock wave velocity in the case of sets 1 and 2 that
involve interactions of clouds with an SSSF, which is in agreement
with the value inferred from the numerical simulations, while for
the simulations of set 3, where the interaction is with a thin SNR
shell, the inferred normalized velocity from the simulations is 100
times smaller than that expected from eq. (14).

The velocity evolution of a cloud depends on the density, mass and size of
the cloud, and on the velocity and density of the shock wave.
We have seen that in the presence of radiative cooling, the cloud
fragmentation and destruction is inhibited and thus we should expect a larger
effective cloud interacting surface with the shock front than in a
non-radiative case which would tend to increase the final cloud velocity.
With the values above for the velocity, we find that the maximum
displacement ($L_M$) that a cloud can undergo before being
destroyed, in units of the cloud radius, is:

\begin{equation}
L_M \sim 2 \psi \ v_{c,N} \ r_c {\rm {(pc)}}
\end{equation}
\noindent
where $\psi$ is:

\begin{equation}
\psi = {{t_{\rm des}} \over {t_{SC}}}
\end{equation}
\noindent
and $t_{\rm des}$ is the destruction time of the cloud.
From the results shown above, we find that in all radiative
cooling simulations $\psi \ge 100$. Thus, the maximum
displacement of the cloud is greater than the value obtained by
PFB02, of the order of 3.5 $r_c$, in their study of SSSF-clouds
interactions, which corresponds to $\psi \sim$ 10.
We can thus conclude that radiative cooling interactions postpone
the destruction of the clouds and increase their velocity. When
shocked by an SNR, eq. (15) indicates that they can be displaced
over distances of $\sim 10-20$ times the cloud radius (for
$v_{c,N} \sim$ 0.5).

\subsection{Mass loading}

When a cloud is shocked either by an SSSF wind or an SNR, its gas is
compressed by a forward shock wave, as discussed in \S2.
After this phase, the compressed gas begins to expand into the ISM and
part of the gas is also ablated by the external shock wave. This
causes an increase of the cloud volume and a decrease of its
density. Theoretically, this phenomenon could be responsible for
the mixing of the cloud gas with the ISM and contribute to
the global increase of the ISM density due to the interaction
between several clouds and SNRs. However, the  simulations above
without photo-evaporation have shown that  the forward shock  that
propagates into the clouds is not able to effectively destroy
them. The situation is apparently different in the case where
photo-evaporation is also considered. In this case, especially in
the runs SNSP1 and SNSP2, the volume filling factor increases to
values greater than 1000 times the initial volume of the cloud.
This means that the gas of the cloud is really mixed with the ISM
and that the mass loading is very efficient. However, 
this result is caused mainly by the
photo-evaporation of the clouds that expand very fast occupying
all the ambient medium. From these simulations, which include the
processes of photo-evaporation, ablation and the passage of the
shock wave through the cloud, we infer that the clouds are
destroyed in at least 20.000 yr, and the corresponding mass loss
rate (in terms of the total cloud mass) is
$\dot m_c \sim 5 \times 10^{-5} m_c {\rm M_{\odot} \ yr^{-1}}$ \footnote{
For strong shocks, the typical timescale for the destruction of the cloud is
mainly dependent on the cloud mass.}.
This is in agreement with the results obtained by Melioli \& de Gouveia Dal
Pino (2004), where it has been shown that the combined effects of
ablation and photo-evaporation of the clouds can lead to an increase
of the ISM density in SB environments.

Now, returning to the cases without photo-evaporation, where we
have only considered the radiative cooling interaction of the
clouds with a SNR (as in the models SNS1 and SNS2), the results
are substantially different. A thin cold dense shell develops
at the contact discontinuity between them which is formed from the
radiative cooling of the cloud shocked material (see \S 3). In
this case, the cloud gas is much less efficiently mixed with the
ISM, and the only important effect resulting from the interaction
is the formation of dense cold filaments with densities up
to $\sim 100$ times larger than that of the ambient medium and
temperatures up to $\sim 50$ times smaller than that of the
ambient medium. These structures are observed in several SBs and
galactic winds (see, e.g., Cecil et al. 2001), and do not
contribute to increase the density of the diffuse ambient medium
(where the observed temperature and density flutuations are,
in general, not greater than a factor $\sim 2$). Comparing our
results with those obtained by PFB02, it is possible to highlight
the fact that in radiative cooling interactions the gas mixing is
reduced and postponed to much later times. In PFB02, after a time
of 68 $t_{SC}$, the gas of each cloud has almost completely mixed
with that of the other clouds and with the ISM; at this time the
clouds have lost their identity and the shock front has destroyed
all the denser structures. In our study, on the other hand, the
clouds become denser, colder filaments and only the
gas of their external parts is dragged by the SNR shell. In the
presence of a flux of UV photons, the photo-evaporation is the
main mechanism responsible for most of the mass loss of the clouds
to the ambient medium, therefore confirming the results of Melioli
\& de Gouveia Dal Pino (2004). This means that radiative cooling
SNR-cloud interactions are unable to destroy the clouds,
especially in an SB environment where there are small, dense clouds
and an ISM with low density. Instead, these interactions produce
cold filaments and do not contribute to significantly increase the
ambient density.

\subsection{Fragmentation and destruction of the clouds}

We determine the lifetime of a shocked cloud and its mass evolution. 
Let us consider the cases of cloud interactions with an SSSF. 
From the study above we can see that the radiative cooling clouds
do not suffer significant fragmentation during the interaction,
and the gas is not efficiently mixed with the rest of the ISM.
The main point is to understand why fragmentation does not
occur. We may distinguish two aspects. First, both the shape and the
mass of the cloud change with time. Second, the perturbations in
the cloud
gas are generally dissipated faster than in the adiabatic case.
The resulting elongated dense filaments still belong to the
original cloud structure and are
therefore separated from the diffuse gas phase of the ISM. 
This result is a consequence of the fact that
all the gas of the cloud is impacted approximately  with the same
intensity and in the same direction (since $r_c < h_{sh}$), and
that the radiative cooling dissipates most of the small
perturbations, preventing their increase with time. 
The shock wave may be unable to destroy the cloud, and other physical 
phenomena, like ablation by the ISM, thermal conduction and 
photo-evaporation, are necessary to have efficient mixing 
(e.g., Melioli \& de Gouveia Dal Pino 2004). 
Certainly, the shocked gas of the cloud tends to
expand because of the increase of its internal pressure, but this
expansion is opposed by the high external pressure of the shocked
ambient medium, as we can see in Figure 18, where the profile of
the pressure of the model SNS1 is shown at a time 81 $t_{SC}$.
The efficient radiative cooling of the shocked gas inside
the cloud inhibits its destruction and mixing with the ISM.

\begin{figure}
\center
\epsfxsize=6cm
\epsfbox{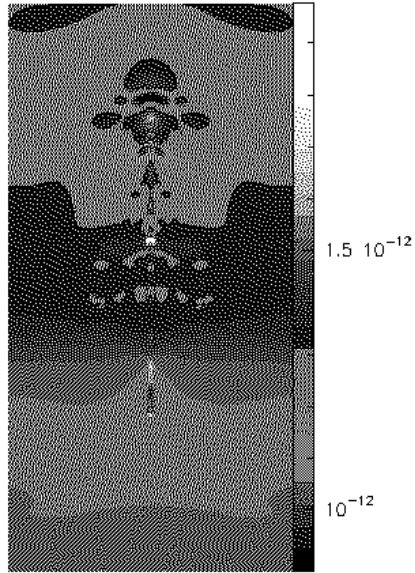}
\caption{Model
SSN1: pressure profile (in log scale) at a time $t=81 t_{sh}$.}
\end{figure}
\noindent
Thus, the interaction between one (or more) cloud(s) and a shock
wave may be considered as a sum of two effects: an ablation caused
by the drag and a re-expansion of the cloud core generated by the
passage of an internal forward shock wave. Taking into account the
typical mass loss rate from drag (Klein, McKee \& Colella
1994), our cloud should be destroyed by the drag in at least
18.000 yr only, which is much shorter than the time we observe in our
simulations, that is $\sim$ 36.000 yr.
This discrepancy increases if we consider that in the study of
PFB02 the clouds are destroyed in $\sim$ 10.000 yr.

The results obtained in this work show that the gas of clouds
is not ablated with the efficiency predicted in previous studies
that did not take into account the effects of radiative cooling
and photo-evaporation, and that SNR-cloud interactions only are
unable to guarantee an efficient mixing of the cloud gas with the
ISM. A single cloud shocked by an SSSF (model SR1) suffers a
considerable lengthening in the direction of the passage of the SSSF and
also in the case of the interaction with a SNR (model SNS1),
and the resulting elongated filamentary cloud
moves without exchanging much mass with the ISM. In the case of
two or three clouds, the mixing is greater, but the elongated
structures are still present. A comparison of these results with
those obtained by PFB02 indicates that the radiative cooling plays
a fundamental role in the evolution of the clouds; they are more
accelerated, only partially fragmented, and its gas is spread in
the ISM with a lower efficiency, so that no global increase of the
ambient density is observed due to these interactions. Only in
the presence of UV photon flux is it possible for the clouds to
undergo an efficient mass loss to the ISM, therefore producing
significant variations in the density, temperature and chemical
composition  of the ISM. In this case the combination of
photo-evaporation, ablation and the shock propagation (which causes
cloud expansion due to heating of the gas) determines the
growth of the ambient density by a factor of 10 with respect to
the initial density.

These simulations have been
performed with initially small clouds, $r_c < h_{sh}$. When the
cloud size is comparable to or larger than the SNR shell width, the
evolution of the cloud is expected to be very different (see,
e.g., Fragile et al. 2004), and instead of gas mixing we may
expect an increase of the cloud density itself over a timescale
sufficiently large to produce cloud collapse and, perhaps, a
process of star formation. This more focused study on processes of
star formation is in progress (Melioli et al., in preparation).

\begin{acknowledgements}
C.M. and E.M.G.D.P acknowledge financial support from the Brazilian Agencies
FAPESP and CNPq.
The authors also acknowledge the careful revision and the useful suggestions 
of the anonymous referee that have helped to improve this work. 
\end{acknowledgements}

{}

\label{lastpage}
\end{document}